\documentstyle[prc,aps,epsfig,preprint]{revtex}

\begin{document}

\draft

\title{Neutral Pions and \protect\bbox{$\eta$} Mesons as Probes of the 
Hadronic Fireball \\ in Nucleus--Nucleus Collisions around 1$A$~GeV}

\author{R.~Averbeck$^{1,*}$, R.~Holzmann$^1$, V.~Metag$^2$, R.~S.~Simon$^1$}
\address{
$^1$Gesellschaft f{\"u}r Schwerionenforschung,
D-64291 Darmstadt, Germany\\
$^2$II. Physikalisches Institut, Universit{\"a}t Gie{\ss}en,
D-35392 Gie{\ss}en, Germany}

\date{\today}
\maketitle

\vspace*{-1cm}
\begin{abstract}
Chemical and thermal freeze--out of the hadronic fireball formed in symmetric 
collisions of light, 
intermediate--mass, and heavy nuclei at beam energies between 0.8$A$~GeV
and 2.0$A$~GeV are discussed in terms of an equilibrated, isospin--symmetric 
ideal hadron gas with grand--canonical baryon--number conservation. 
For each collision system the baryochemical potential 
$\mu_B$ and the chemical freeze--out temperature $T_c$ are deduced 
from the inclusive $\pi^0$ and $\eta$ yields which are augmented by 
interpolated data on deuteron production. With increasing beam energy $\mu_B$ 
drops from 800~MeV to 650~MeV, while $T_c$ rises from 55~MeV to 90~MeV.
For given beam energy $\mu_B$ grows with system size, whereas
$T_c$ remains constant.
The centrality dependence of the freeze--out parameters is weak as 
exemplified by the system Au~+~Au at 0.8$A$~GeV.
For the highest beam energies the fraction of nucleons excited to resonance 
states reaches freeze--out values of nearly 
15\%, suggesting resonance densities close to normal nuclear density 
at maximum compression.
In contrast to the particle yields, which convey the status at 
chemical freeze--out, the shapes of  
the related transverse--mass spectra do reflect thermal 
freeze--out. The observed thermal freeze--out temperatures $T_{th}$
are equal to or slightly lower than $T_c$, indicative of nearly simultaneous
chemical and thermal freeze--out. 
\end{abstract}


\narrowtext

\section{Introduction}

Relativistic nucleus--nucleus collisions offer the unique possibility to
study nuclear matter under the influence of high temperature and pressure.
At incident energies around 1$A$~GeV the formation of a hot and dense 
reaction zone, dubbed the fireball, has been verified 
experimentally \cite{STO86}.
According to model calculations, the compression phase lasts for time
spans of 10~--~15~fm/c and reaches densities of 2~--~3 times
the nuclear ground--state density \cite{CAS90,AIC91,MAR94,BAS95}.
Simultaneously, temperatures up to $\approx$100~MeV may be achieved and a 
substantial fraction of the nucleons participating in the collision is 
excited to heavier, short--lived resonance states which decay predominantly 
via meson emission \cite{EHE93,MET93,MOS93}.
Thus the fireball produced in the energy regime of the
heavy--ion synchrotron SIS at GSI Darmstadt comprises nucleons, resonances, 
and mesons.

It is an interesting question to what extent this hadronic system can be 
described in terms of chemical and thermal equilibrium.
In the present paper we address this issue on the basis of inclusive 
as well as centrality--selected data on 
$\pi^0$ and $\eta$--meson production in collisions of nuclei with
equal mass number $A$.
Chapter~2 introduces the relevant concepts and gives a brief account 
of previous studies of the hadronic fireball at SIS energies.
The existing systematics of $\pi^0$ and $\eta$ production is 
reviewed in chapter~3. There we demonstrate that
thermal concepts do provide a useful description of the fireball 
at SIS energies.
Our model of an ideal hadron gas in chemical equilibrium is 
presented in chapter~4.
Chapter~5 discusses the results of the thermal analysis. 
The chemical composition of the fireball at freeze--out is 
determined, and the consistency of chemical and thermal freeze--out 
temperatures is addressed. Finally, the freeze--out conditions of hadronic 
matter as derived for the SIS energy regime are compared to 
results obtained from similar analyses of particle ratios measured at 
significantly higher incident energies at the 
AGS (Brookhaven National Laboratory) and 
the SPS (CERN).

\section{Thermal Concepts in Nucleus--Nucleus Collisions at SIS
Energies}

In the initial phase of nucleus--nucleus collisions with incident
kinetic energies $E_{beam}$ around 1$A$~GeV a 
system of interacting nucleons, resonances, and mesons is created.
The size of the hadronic fireball depends on the masses of target and projectile
nucleus and, in addition, on the centrality of the collision.
Within the participant--spectator model, the geometrical overlap of the 
two nuclei determines the number of nucleons $A_{part}$ which are directly 
involved. The quantity B = $A_{part}$ thus defines the number of baryons 
in the system. For symmetric collisions and complete stopping 
the energy available per baryon is $\sqrt{s}/2-m_N$, with s given by 
$4m_{N}^{2}+2m_{N}E_{beam}$ and $m_{N}$ denoting the nucleon mass.
The initial conditions of the fireball are therefore fixed.
In the subsequent time evolution the available energy is 
transformed into the excitation of thermal and collective degrees of 
freedom. The energy turns into heat and provides the mass stored in 
resonance states and mesons at chemical freeze--out.
In addition, the energy builds up compression and produces the flow of the 
expanding matter.

In a simplified picture hadrons cannot escape from the fireball during
the high--density phase of the collision.
Nucleons, resonances, and mesons are trapped in a cyclic process of 
generation, absorption, and re--emission, exemplified for nucleons $N$, 
$\pi$ mesons, and $\Delta$ resonances by 
$NN \rightleftharpoons N\Delta \rightleftharpoons NN\pi$.
Within this approach hadrons are released only with the onset of the 
expansion phase, when mesons and baryons decouple
due to the decreasing matter density. 
Moments in the expansion process when certain degrees of freedom of the 
system no longer participate in the interaction provide landmarks in the 
time evolution.
One has to distinguish between chemical and thermal freeze--out which
-- in the limit of sudden freeze--out -- 
correspond to those moments in time when the relative abundances of the 
particle species or their momentum distributions stop to change. 
While only inelastic collisions involving the short--range
nuclear force can alter the relative 
particle yields, the momentum distributions of the particles 
are governed by the larger total interaction cross sections.
Consequently, thermal freeze--out does not occur before chemical
freeze--out.
Through the frequent scattering processes of the constituents 
the system maintains chemical and eventually thermal equilibrium.

It has been shown that at the AGS ($E_{beam} \leq$ 13.7$A$~GeV) and the 
SPS (200$A$~GeV) a large number of hadronic 
observables, including strange--particle yields, is in quite good agreement 
with such an equilibrium scenario \cite{CLE93,BRA95,BRA96,SOL97}.
At the much lower SIS energies, however, only a very limited 
variety of hadron species is produced with significant yields.
Therefore, in contrast to the situation at AGS and SPS energies, 
the number of observables which may reflect 
the chemical or thermal freeze--out is comparable to 
the number of free parameters in common thermal models. 
Nevertheless, several analyses have been performed to check for
consistency between data and thermal--model predictions also at 
SIS energies.
Midrapidity transverse--momentum spectra of charged pions, protons, and 
deuterons measured in central Ni~+~Ni collisions at 1.06$A$, 1.45$A$, and 
1.93$A$~GeV 
were observed to be consistent with thermal equilibrium 
if, in addition, collective radial flow was taken into account \cite{HON98}.
For the same reactions also chemical equilibrium has been
claimed with close agreement between the chemical freeze--out temperatures
and the temperatures derived from the particle spectra.
Recently, Cleymans et al.~have extended the thermal analysis at 
SIS energies to comprise also the strange mesons K$^+$ and K$^-$. 
In their systematic study of central collisions of Ni~+~Ni and Au~+~Au  
based on a model 
with canonical strangeness conservation \cite{CLE98,CLE99} 
the yields of protons, deuterons, charged pions, and K mesons were found to
agree with chemical equilibrium at freeze--out. Failure of 
their analysis to also accomodate the $\eta$ meson
within this common freeze--out picture \cite{CLE99} may be attributed 
to the use of extrapolated $\eta$ yields.

\section{Evidence for thermal behavior from meson yields and spectra}

Pions and $\eta$ mesons are the most abundantly produced mesons
at SIS energies.
While the pionic degree of freedom is also covered by the spectroscopy of 
the charged members of the isospin multiplet, it is only through 
$\gamma$-ray spectroscopy that the $\eta$ meson becomes observable in 
nucleus--nucleus collisions.
The two neutral mesons $\pi^0$ and $\eta$ can both be  
identified by a two-photon 
invariant--mass analysis of coincident photon pairs. Using the Two--Arm 
Photon Spectrometer, the TAPS collaboration performed a series of 
systematic meson--production experiments covering incident energies from 
0.2$A$~GeV to 2.0$A$~GeV. 
These measurements have established an extended data base for $\pi^0$ and 
$\eta$ production in light (C~+~C), intermediate--mass (Ar, Ca~+~Ca), 
and heavy symmetric systems (Ni~+~Ni, Kr~+~Zr, Au~+~Au) 
\cite{WOL98,BER94,SCH94,APP97,AVE97,MAR97,MAR98,VOG98}.

\begin{figure}
\begin{center}
\mbox{\epsfig{file=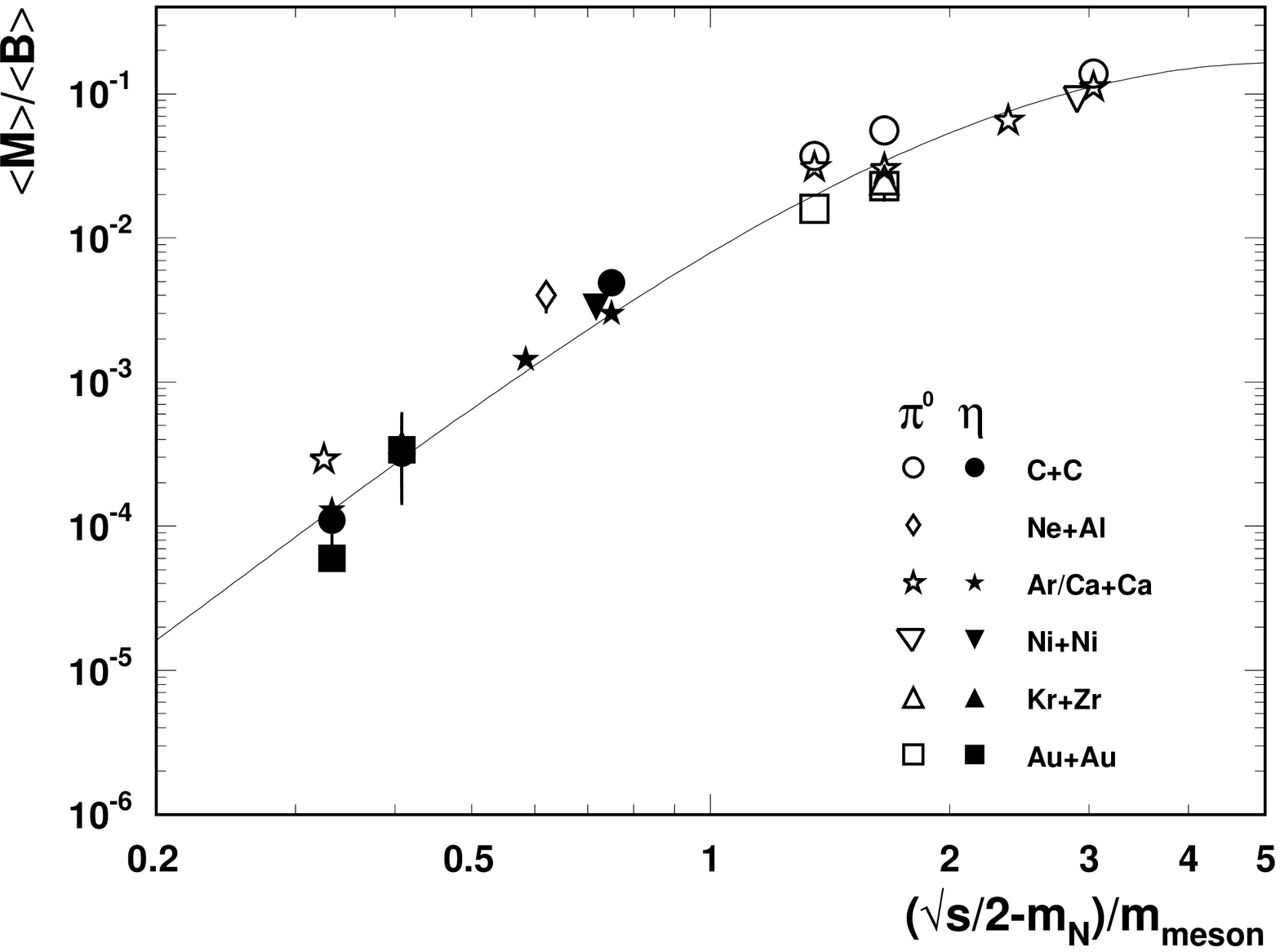,width=\columnwidth}}
\caption[]{Average $\pi^0$ and $\eta$--meson multiplicities per average
number of participants in symmetric nucleus--nucleus collisions as a function of
the energy available per baryon normalized to the mass 
of the respective meson. The data are taken from
\cite{WOL98,BER94,SCH94,APP97,AVE97,MAR97,MAR98,VOG98}. The 
curve represents a fit to the data and is given by the polynomial expression
\begin{math}
\log(\langle M \rangle / \langle B \rangle) = 
-2.102 + (3.25 - (1.405 + 0.785 \log x) \log x) \log x ,    
\end{math}
where $x$ is the normalized meson--specific energy available.
}
\label{metag}
\end{center}
\end{figure}

The primary information for the present discussion is provided by 
the inclusive meson yields.
They have been determined within narrow rapidity intervals around 
midrapidity. For a consistent treatment the data of the various systems   
are extrapolated to the full solid angle assuming an isotropic  
source at midrapidity.
Fig.~\ref{metag} shows the resulting average inclusive meson multiplicities 
$\langle M \rangle$, normalized to the 
average number of participating baryons $\langle B \rangle$, as a 
function of the energy available per baryon 
divided by the corresponding meson mass.
Except for the system Au~+~Au at 0.8$A$~GeV \cite{WOL98} and the 
systems Au~+~Au and Kr~+~Zr at 1.0$A$~GeV (see table \ref{data}), where
$\langle B \rangle$ has been determined experimentally, 
we calculate $\langle B \rangle$ from the geometrical overlap of two 
colliding sharp spheres which gives $\langle B \rangle = A/2$ in 
case of collision partners with equal mass number $A$.
A steep rise of the normalized meson ratios with increasing energy is observed.
In addition, a weaker dependence on the size of the collision system is 
visible with the clear tendency towards smaller inclusive yields in the 
heavier systems. 
The same trends have also been observed in charged--pion production experiments
\cite{MUE97,PEL97}.
In first approximation all data points fall onto a smooth curve 
indicating that to a large extent the meson--production probability is 
determined by the energy available per baryon. 
This is quite remarkable because $\pi^0$ and $\eta$  production proceed 
through different baryon resonances: $\pi^0$ mesons mainly come from 
$\Delta(1232)$--resonance decays, while the heavier $\eta$ mesons essentially 
originate from the $N(1535)$ resonance which, at SIS energies, is the only
significantly populated baryon resonance with a large decay
width into $\eta$ mesons.
The fact that the basic production mechanism is no longer apparent in  
the observed meson yields may be interpreted as a first
indication for meson emission from an equilibrated source.

\begin{figure}
\begin{center}
\mbox{\epsfig{file=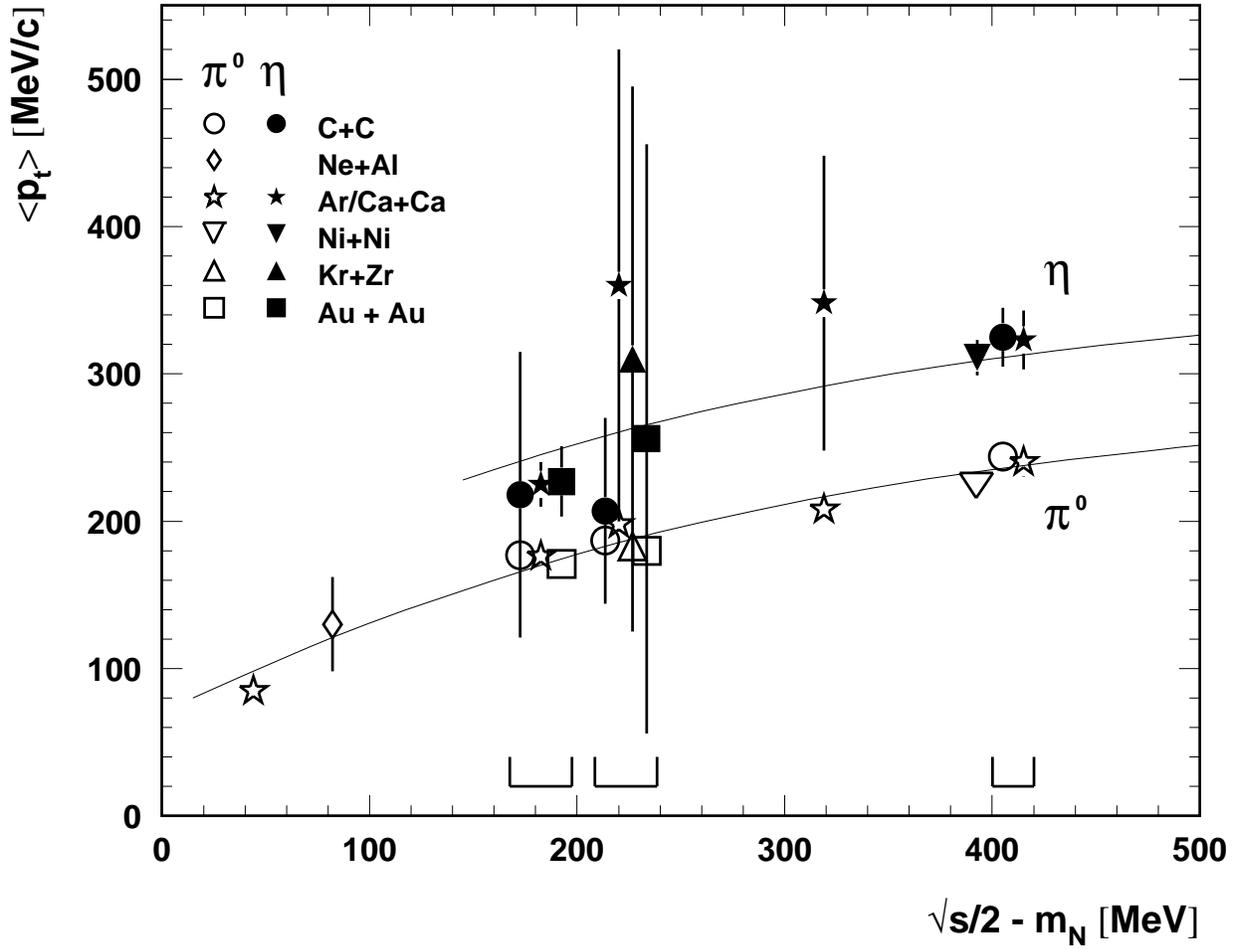,width=\columnwidth}}
\caption[]{Average $\pi^0$ and $\eta$--meson transverse momenta measured
in narrow rapidity intervals around midrapidity as a function of the energy
available per baryon. The data are taken from the same
experiments as listed in Fig.~\ref{metag}. The data points at available
energies of 182~MeV (0.8$A$~GeV beam energy), 223~MeV (1.0$A$~GeV), and
410~MeV (2.0$A$~GeV) are slightly shifted in energy with respect to 
their nominal position (center of brackets) to make the error bars of the
$\eta$ momentum visible. With the exception of Ne~+~Al the error bars for the 
$\pi^0$ momentum are smaller than the symbol size.
Parallel lines interpolating 
the $\pi^0$ and $\eta$ data, respectively, are drawn to guide the eye. 
}
\label{ptsys}
\end{center}
\end{figure}

The second important result concerns the average transverse momenta 
$\langle p_t \rangle$ at midrapidity. 
For the decay of nominal--mass
$\Delta(1232)$ and $N(1535)$ resonances 
one has center--of--mass momenta of 229 MeV/c for $\pi^0$
from $\Delta(1232)$ and 182 MeV/c for $\eta$ from $N(1535)$.  
For resonances embedded in an isotropic fireball one therefore 
expects average transverse momenta obeying 
$\langle p_t \rangle_{\pi^0} > \langle p_t \rangle_{\eta}$.
Fig.~\ref{ptsys} shows the experimental results 
as a function of the energy available per baryon.
Although the uncertainties for the $\eta$ data are large 
compared to those of the $\pi^0$ data, the $\eta$ momenta are clearly 
higher than the $\pi^0$ values, opposite to the expectation.
For both mesons the initial rise of $\langle p_t \rangle$ 
slows down with increasing available energy, indicating
possible saturation of the average transverse momentum.
We conclude that also with respect to the shapes of the meson spectra the
naive expectations based on the production mechanism are not borne out 
experimentally.

As a third piece of evidence for thermal behavior we mention the 
transverse--mass spectra of the mesons.
At midrapidity the transverse mass $m_t$ of a particle is equivalent to its
total energy in the center--of--mass system.
Following \cite{HAG83,STA92}, the transverse--mass distribution of particles 
with mass $m$ emitted isotropically from a thermal source 
is characterized by a Boltzmann temperature 
$T_B$ and, at midrapidity, can be approximately described by

\begin{equation}
\frac{1}{m_t^2}\,\frac{d\sigma}{dm_t} \propto 
\exp\left(-\frac{m_t}{T_B}\right) 
\;\;\mbox{with}\;\; m_t = \sqrt{m^2 + p_t^2}~.
\label{mtdis}
\end{equation}

\noindent
In Fig.~\ref{mtscal} transverse--mass distributions of $\pi^0$ and $\eta$
mesons are plotted together with fits according to the parametrization 
given in Eq.~\ref{mtdis} for three different systems at incident energies 
near 2$A$~GeV.
Within each reaction the $\pi^0$ and $\eta$ spectra exhibit almost identical
inverse--slope parameters $T_B$ which in addition do not change significantly
with the mass of the colliding nuclei.
In all three systems the $\pi^0$ and $\eta$  
intensities roughly coincide for $m_t \ge m_\eta$.
This indicates that it is the energy required to produce a given transverse 
mass which determines the relative abundance of the meson species near 
midrapidity. For low $m_t$, however, 
individual differences among the collision systems become apparent.
Fig.~\ref{mtscal} shows a systematic enhancement over the exponential rise 
extrapolated from the high--$m_t$ region if one goes from the light C~+~C to
the heavy Ni~+~Ni system.
The same observations, namely $m_t$ scaling of the $\pi^0$ and $\eta$ 
intensities 
and low--$m_t$ enhancement of the $\pi^0$ spectrum in heavy systems, 
have been reported at various energies down to 
0.8$A$~GeV and seem to be a general feature of heavy--ion collisions
in the SIS--energy regime \cite{WOL98,BER94,APP97,AVE97,MAR97,VOG98}.
Possible explanations that have been suggested for the 
low--$m_t$ enhancement involve pion rescattering 
through resonance states in 
the heavier systems \cite{BAS93} and multiple pion decay of heavy 
resonances \cite{TEI97}.

Transverse--mass scaling in the high--$m_t$ region does not prove that the 
fireball actually has reached chemical equilibrium. The particle spectra
of an equilibrated source, however, would follow phase space distributions as 
observed, at least for vanishing flow.

\section{Ideal Hadron--Gas Model} 

In our ansatz we assume that at chemical freeze--out the
fireball can be described in terms of an ideal, equilibrated hadron gas.
As constituents we take into account pions and $\eta$ mesons, as well as 
nucleons, deuterons, and all non--strange baryon resonances up to a mass of 
1.8~GeV (see table \ref{resonances}), which corresponds to $\sqrt{s}-m_N$ 
at 2$A$~GeV beam energy.

In the grand--canonical description chosen, hadronic matter 
is characterized by a baryo\-chemical potential $\mu_B$ and a 
temperature $T_c$. Furthermore, one might consider the isospin of the system. 
In the present analysis, however, we neglect 
isospin as an additional degree of freedom, mainly because
our data base of $\pi^0$ and $\eta$ yields is insensitive to isospin.
Within the isobar model for instance, the $\pi^0$ yield amounts to 
one third of the total pion yield of a heavy--ion collision irrespective 
of the actual isospin of the projectile--target system.
The other important observable at SIS energies, the $\eta$ meson, does not 
carry isospin.
Thus, only the third observable in our data base, the relative yield 
of deuterons and nucleons, depends on the isospin.

In the grand--canonical description of a system of noninteracting
fermions and bosons the particle--number densities $\rho_i$ are 
given by integrals over the particle momentum $p$

\begin{equation} 
\rho_{i} = \frac{g_i}{2\pi^2}\;
           \int_0^\infty\frac{p^2\,dp}
           {exp\left[\left(E_{i}-\mu_{B} B_{i} \right)/T_c \right] \pm 1}~,
\label{grandcan}
\end{equation} 

\noindent
where $g_i$ is the spin--isospin degeneracy, $E_i$ the total energy in the
local restframe, and $B_i$ the baryon number of the particle species $i$.
The form of the denominator in the integrand accounts for the different 
statistics of fermions and bosons.

Eq.~\ref{grandcan} describes an infinitely large system of stable particles
and  
cannot be directly applied to the hadronic fireball in a nucleus--nucleus
collision. 
To account for the fact that the fireball occupies only a finite volume,
a surface correction has to be included.
We assume a spherical freeze--out volume $V_c$ of
radius $R_c$ and obtain a momentum--dependent correction factor $f(pR_c)$ 
\cite{BAL70,JAQ84} 

\begin{equation}
f(pR_c) = 1 - \frac{3\pi}{4pR_c} + \frac{1}{\left(pR_c\right)^2}
\label{surface}
\end{equation} 

\noindent
in the integrand of Eq.~\ref{grandcan}.
The correction leads to a 30--40\% reduction of the individual
particle--number densities as compared to infinite nuclear matter. 
Particle ratios, however, are hardly affected since here 
the corrections nearly cancel. 
The freeze--out radius of the system is fixed by baryon number conservation
\begin{equation} 
\int_{V_c} \rho_B dV  =\langle B \rangle ~,
\label{radiusfix}
\end{equation}
where the baryon density $\rho_B$ comprises contributions from
nucleons, deuterons, and resonances
\begin{equation} 
\rho_B = \rho_N + 2\rho_d + \sum\limits_{R}\rho_{R} ~.
\label{baryondense}
\end{equation}
Given the feedback via Eq.~\ref{surface}, 
$R_c$ has to be adjusted iteratively. Starting from an ad--hoc initial value,
$R_c$ is varied in steps of
0.5~fm in order to fulfill Eq.~\ref{radiusfix} within the  
uncertainties of the experimental particle ratios.
 
The second modification of Eq.~\ref{grandcan} is related to the fact 
that the baryon resonances are unstable particles which do not have a fixed 
mass but exhibit a broad mass distribution.
In contrast to thermal--model analyses of AGS and SPS data, the actual
mass distributions cannot be neglected in the 1$A$~GeV
energy regime since here the energy available in the 
nucleon--nucleon system is comparable to the excitation energy
of the baryon resonances.
This means that the $\Delta(1232)$ resonance, the lowest of the 
baryon resonances, is by far the most abundantly 
populated resonance and that all resonances are populated 
predominantly in the low--mass tail of their mass distributions. 
To account for the off--shell behavior of the resonances, the integrand in 
Eq.~\ref{grandcan} is folded with normalized mass 
distributions $A_i(m)$. These are trivial delta functions for the stable 
nucleon and deuteron and for the electromagnetically decaying $\pi^0$
and $\eta$, while the scheme to parametrize the resonance mass distributions 
has been adopted from \cite{TEI97}.

We describe a given  resonance mass distribution by a Lorentzian 
of the form
\begin{equation}
A_R(m) \propto \frac{m^2\Gamma_{R}(m)}
              {\left(m^2-m_R^2\right)^2 + m^2\Gamma_{R}^2(m)}~, 
\label{lorentzian}
\end{equation}
\noindent
with
\begin{equation}
\int_0^\infty A_R(m)\,dm = 1~. 
\label{anorm}
\end{equation}

\noindent
The characteristic parameters are the nominal resonance mass 
$m_R$ and the mass--dependent total decay width $\Gamma_{R}(m)$.
As decay modes we take 1$\pi$, 2$\pi$, and $\eta$ decay to the nucleon 
ground state
into account. The total decay width at mass m thus is given by
\begin{equation}
\Gamma_{R}(m)=\Gamma_{1\pi}(m)+\Gamma_{2\pi}(m)+\Gamma_{\eta}(m)~.
\end{equation}

\noindent
The 2$\pi$ decay is described
in terms of a one--step process where the resonance decays into
a nucleon and an object with angular momentum $l=0$ and twice the pion mass 
m$_{2\pi}$ = 2m$_\pi$. The hypothetical di--pion subsequently 
disintegrates into two pions. With this simplification the 2$\pi$
decay width can be written in complete analogy to the
1$\pi$ and $\eta$ widths.
For the mass--dependent partial decay widths 
$\Gamma_{1\pi,2\pi,\eta}$(m) we thus obtain  
      \begin{equation}
      \Gamma_{1\pi,2\pi,\eta}(m) = 
            \left\{
            \begin{array}{l@{\quad\quad\quad}l@{\quad}l}
            \displaystyle
            \Gamma_{1\pi,2\pi,\eta}(m_R) 
            \left(\frac{q}{q_R}\right)^{2l+1}
            \left(\frac{q_R^2+\delta^2}{q^2+\delta^2}\right)^{l+1}  
            & m > m_N + m_{\pi,2\pi,\eta}
            & ,\\
            0 
            & \mbox{otherwise}
            & .
            \end{array}
            \right. 
\label{partialwidth}
      \end{equation}
\noindent
The partial decay widths at 
m = $m_R$ are listed in Tab.~\ref{resonances}. The quantity $l$ is the 
angular momentum of the emitted meson or di--pion, and $q$ is the momentum
of the particle in the restframe of the decaying resonance, with $q_R$ denoting
the special case of m = m$_R$. The cutoff 
parameter $\delta$ is given by
      \begin{equation}
      \delta^2 = \left(m_R-m_N-m_{\pi,2\pi,\eta}\right)^2 
                 + \frac{\Gamma_R^2(m_R)}{4}~.
      \end{equation}
\noindent
Only in the case of the $\Delta$(1232) resonance we deviate from 
this prescription, mainly for the sake of consistency. Use
of the well--established parameters $\Gamma_R$ = 110 MeV and 
$\delta$ = 300 MeV/c
determined for this resonance by Koch et al.~in a non--relativistic 
analysis (see \cite{KOC84}) requires
an additional term m$_R$/m on the right--hand side of 
Eq.~\ref{partialwidth}.  

Furthermore, one could consider an excluded--volume correction to take into
account the hadron--hadron hard--core repulsion. This means that one would 
consider a real rather than an ideal hadron gas. An excluded--volume 
correction, however, does not play a significant role for particle--yield
ratios as demonstrated in \cite{SOL97}.
Another reason to omit the excluded--volume correction is the fact 
that chemical freeze--out seems to occur at baryon densities well below
the nuclear ground--state density.
The system therefore is rather dilute which supports the assumption
of an ideal gas of pointlike non--interacting particles.

With the modifications introduced by Eqs.~\ref{surface} and~\ref{lorentzian} 
the particle--number
densities of Eq.~\ref{grandcan} are given by

\begin{eqnarray} 
& \rho_{i} \; & = \frac{g_i}{2\pi^2} \;
           \int_0^\infty dp\,p^2 f(pR_c) \nonumber \\
& &        \int_0^\infty dm\,\frac{A_i(m)}
           {exp\left[\left(\sqrt{m^2+p^2}-\mu_{B} 
           B_{i} \right)/T_c \right] \pm 1}~. 
\end{eqnarray} 

\noindent
Equation \ref{lorentzian} employs a mass-dependent width in the
resonance-mass distribution.
This feature of the model makes it possible
to take the actual resonance masses into account
if one determines the contributions to the 
asymptotically observed $\pi^0$ and $\eta$ intensities 
which originate from resonance decay. 
The effective branching ratios may deviate significantly from
those at the nominal mass m$_R$. For the latent meson densities 
$\rho^{1\pi,2\pi,\eta}_{R}$ 
represented by a given resonance R one obtains
\newpage
\begin{eqnarray} 
& \rho^{1\pi,2\pi,\eta}_{R} \; & = w\;\frac{g_R}{2\pi^2}\;
           \int_0^\infty dp\,p^2 f(pR_c)  \nonumber \\
& &        \int_0^\infty dm\,
           \frac{\Gamma_{(1\pi,2\pi,\eta)}(m) A_R(m)}
           {\Gamma_{R}(m) exp\left[\left(\sqrt{m^2+p^2}-\mu_{B} 
           \right)/T_c \right] + 1} \;\;,
\label{resdecay}
\end{eqnarray} 

\noindent
with $w=1$ in the case of one--pion or $\eta$--meson emission and with $w=2$
for the two--pion decay. The resonances present at freeze--out 
also contribute to the asymptotically observed 
abundance of nucleons. Furthermore, the  
$\eta$ meson (lifetime c$\tau$ = $1.7 \times 10^5$ fm) gives rise  
to additional $\pi^0$ intensity via its 
3$\pi^0$ and $\pi^+\pi^-\pi^0$ decays with branching
ratios of 32\% and 23\%, respectively.   

\section{Thermal--Model Analysis}

The values for the two parameters of the hadron--gas model -- the 
baryochemical potential $\mu_B$ and the temperature $T_c$ -- can be 
derived from any two relative abundances of the constituents. Furthermore,
the baryon content of the system has to be known for an
assessment of the geometrical size of the fireball.
The $\pi^0$ and $\eta$--meson yields are particularly suited for such an 
analysis, as these particles unambiguously arise from the fireball.
Hence, we base our analysis on the ratios 
$\langle M_{\pi^0} \rangle /\langle B \rangle$ and 
$\langle M_{\eta} \rangle/ \langle M_{\pi^0} \rangle$ as compiled in 
Tabs.~\ref{data} and \ref{aucendeptab}.
For the sake of redundancy in the analysis
we consider the ratio of deuterons to nucleons as additional input information. 
The knowledge of $\langle M_{\rm{d}} \rangle /\langle M_{\rm{N}} \rangle$,
however, is limited and especially its centrality 
dependence is difficult to determine experimentally since in 
non--central collisions it is not trivial to separate deuterons originating
from the fireball from those which are emitted by target-- or 
projectile--like spectator remnants.

So far, only a few investigations have addressed 
deuteron yields in nucleus--nucleus collisions at SIS energies.
The inclusive values for 
$\langle M_{\rm{d}} \rangle /\langle M_{\rm{N}} \rangle$
given in Tab.~\ref{data} are based on yield ratios of 
deuterons and protons measured in central collisions of Ni~+~Ni at 
1.06, 1.45, and 1.93$A$~GeV \cite{HON98} and of Au~+~Au 
at 1.0$A$~GeV \cite{POE93}, and on inclusive results obtained for various 
combinations of light and intermediate--mass nuclei at 0.8$A$~GeV \cite{NAG81}.
>From these data we parametrized the dependence of $\langle M_{\rm{d}} \rangle 
/\langle M_{\rm{N}} \rangle$ on the incident energy and on the system
size. For given system size a linear decrease of 
$\langle M_{\rm{d}} \rangle /\langle M_{\rm{N}} \rangle$ with 
increasing available energy provides a fair description of the 
existing data. For fixed energy the ratio $\langle M_{\rm{d}} \rangle 
/\langle M_{\rm{N}} \rangle$ increases with system size
until saturation sets in for heavy systems. 
The limited number of data points and the uncertainties 
in the interpolation procedure translate into quite large errors
for the ratios $\langle M_{\rm{d}} \rangle /\langle M_{\rm{N}} \rangle$
(see Tab.~\ref{data}). No attempt has been 
made to establish an impact--parameter dependence.

In the thermal analysis the three particle ratios are related to the 
particle--number densities 
calculated within the thermal model by the following equations:

\begin{eqnarray}
\left(\frac{\langle M_{\pi^0} \rangle }{\langle B \rangle 
}\right)_{exp} & = &\quad 
          \frac{\displaystyle 1}{\displaystyle \rho_B}\;
          \Bigg\{ 
          \frac{\displaystyle 1}{\displaystyle 3}\,
          \Big( \rho_\pi + \sum\limits_{R}\rho_{R}^{1\pi} + \rho_{R}^{2\pi}
          \Big)
          \nonumber\\
               &   &\quad\quad\quad\quad
          +\quad 1.2\, \Big( \rho_\eta + \rho_{N(1535)}^\eta
          \Big)
          \Bigg\}  ~, \\
& & \nonumber\\
\left(\frac{\langle M_{\eta} \rangle }{\langle M_{\pi^0} \rangle 
}\right)_{exp} & = &\quad 
          \frac{\displaystyle\rho_\eta + \rho_{N(1535)}^\eta}
          {
          \frac{\displaystyle 1}{\displaystyle 3}\,
          \Big(
          \displaystyle\rho_\pi + \sum\limits_{R}\rho_{R}^{1\pi} + 
          \rho_{R}^{2\pi}
          \Big)
          + 1.2\, \Big(
          \displaystyle\rho_\eta + \rho_{N(1535)}^\eta
          \Big)
          }~, \\
& & \nonumber\\
\left(\frac{\langle M_{\rm{d}} \rangle }{\langle M_{\rm{N}} \rangle 
}\right)_{exp} & = &\quad 
         \frac{\displaystyle\rho_{\rm{d}}} 
         {\displaystyle\rho_N + \sum\limits_{R}\rho_{R}}~. 
\end{eqnarray}
\noindent
Each equation defines a band in the 
$\mu_B$,$T_c$--diagram. The widths of these bands are given 
by the uncertainty of the input values, and
for each individual collision system which is characterized by its triplet of 
particle ratios one obtains the chemical potential and the temperature as 
coordinates of the common intersection point of the three bands.
Provided the experimental uncertainties correspond to Gaussian errors 
one has a probability of 20\% that the true values of the three particle 
ratios simultaneously fall into the 1$\sigma$ uncertainty ellipsoid,
while this probability rises to 75\% for 2$\sigma$ uncertainty \cite{PDGSTAT}.
For a consistent treatment of the full data sample it is therefore necessary 
to allow for uncertainties in excess of two standard deviations in the 
input ratios.

The system size also has to be considered as input information.
Through baryon conservation expressed in the grand--canonical form 
of Eq.~\ref{radiusfix} the baryon number determines the
freeze--out radius $R_c$.
For typical values of $R_c$ around 5 fm we find that changes within 
$\pm$1~fm of the final solution hardly affect the numerical values for
$\mu_B$ and $T_c$, but substantially alter 
the freeze--out density $\rho_B$ due to its 1/$R^3_c$ dependence. 
Larger variations up to $\pm$2~fm may still lead to solutions on the 
2$\sigma$--level for the particle ratios
with changes in $\mu_B$ and $T_c$ within $\pm$5\%, but $\rho_B$ becomes 
increasingly incompatible with Eq.~\ref{radiusfix}.

\begin{figure}
\begin{center}
\mbox{\epsfig{file=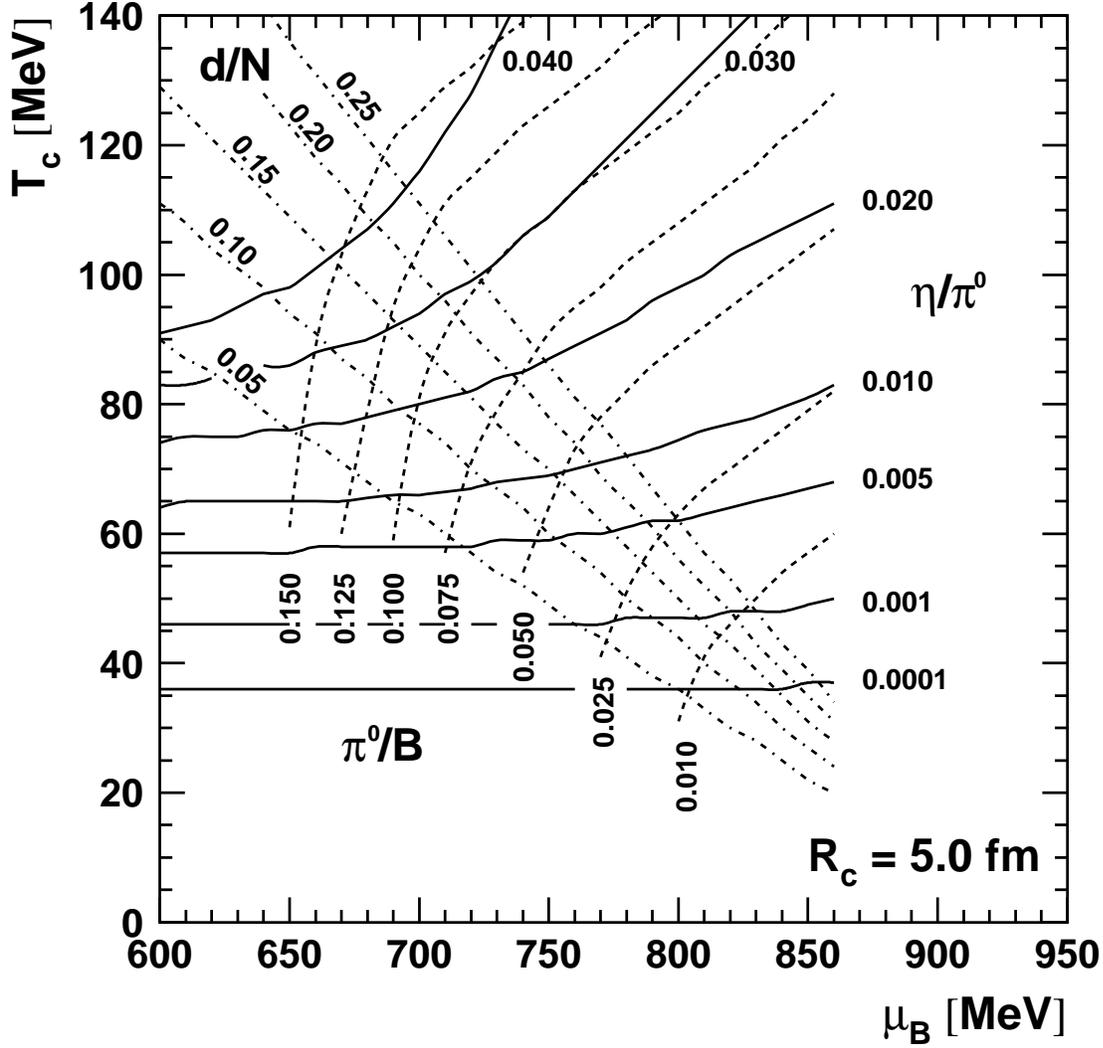,width=\columnwidth}}
\caption[]{
Freeze--out lines in the $\mu_B,T_c$--plane for selected values of
the particle ratios $\pi^0 / \rm{B}$, 
$\eta / \pi^0$, and $\rm{d} / \rm{N}$.
The diagram corresponds to a fixed size of the system with the consequence that
the baryon density $\rho_B(\mu_B,T_c)$ changes. Representative densities are
$\rho_B(650~MeV,70~MeV)$=0.053 $\rho_0$, 
$\rho_B(650~MeV,100~MeV)$=0.469 $\rho_0$, 
$\rho_B(800~MeV,80~MeV)$=1.197 $\rho_0$, and 
$\rho_B(800~MeV,50~MeV)$=0.125 $\rho_0$, 
where $\rho_0$ is the nuclear
ground--state density $\rho_0$ = 0.168~fm$^{-3}$. 
}
\label{mutfixradius}
\end{center}
\end{figure}

\begin{figure}
\begin{center}
\mbox{\epsfig{file=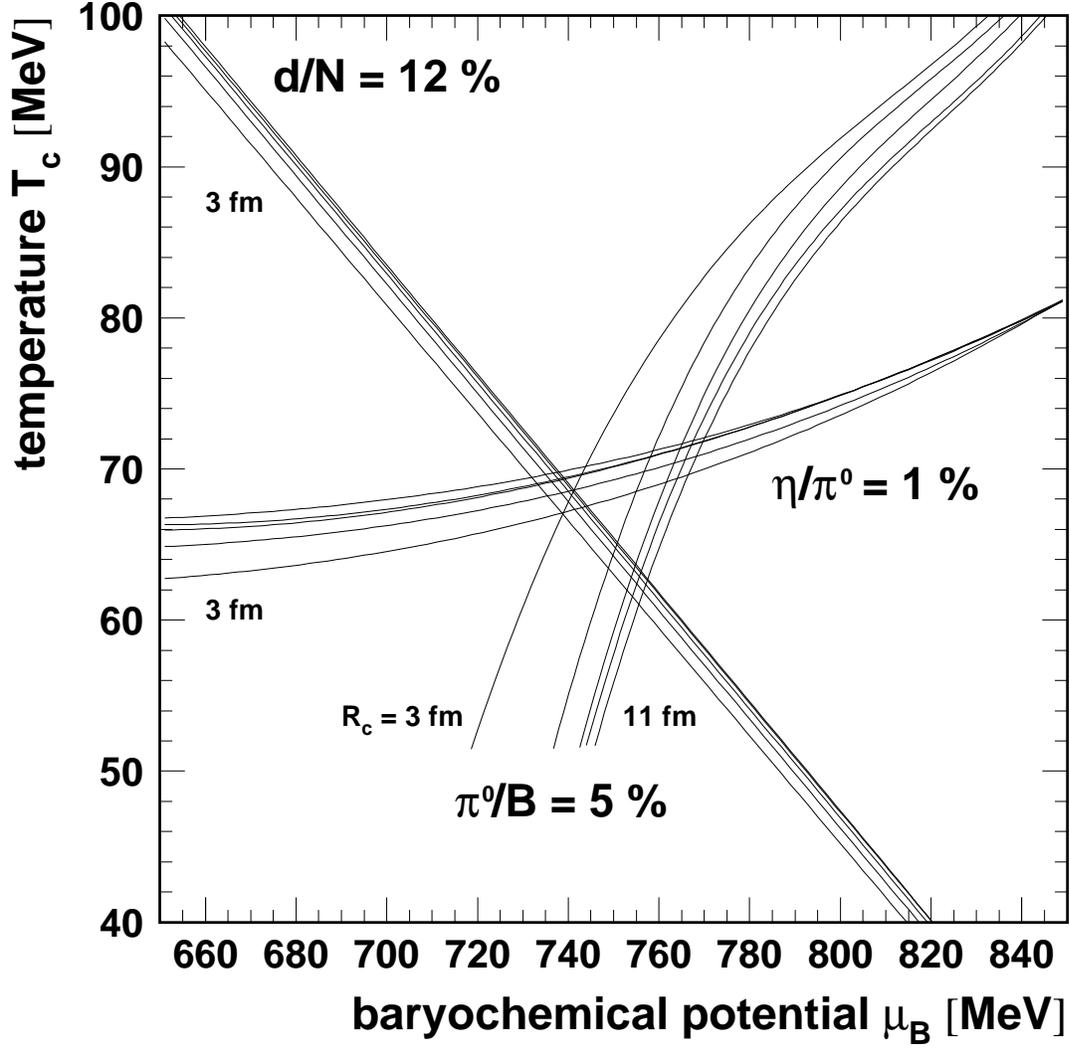,width=\columnwidth}}
\caption[]{
Freeze--out lines in the $\mu_B,T_c$--plane for selected values of
the freeze--out radius $R_c$.
The diagram corresponds to fixed 
particle ratios $\pi^0 / \rm{B}$, 
$\eta / \pi^0$, and $\rm{d} / \rm{N}$, while $R_c$ varies in equidistant 
steps from 3 fm to 11 fm as indicated. 
}
\label{mutfixratio}
\end{center}
\end{figure}

The connection between a given particle ratio and the freeze--out
parameters $\mu_B$ and $T_c$ is illustrated in Fig.~\ref{mutfixradius}. A 
global value of 5 fm for the radius parameter $R_c$ has been chosen to 
generate the diagram. Curves of constant 
particle ratio in the $\mu_B,T_c$--plane are referred to as freeze--out lines.  
Since $R_c$ is fixed, $\langle B \rangle$ varies along the freeze--out lines.
The intersection of any two freeze--out lines from different particle ratios 
thus only describes a physical solution if also the corresponding 
values of $\langle B \rangle$ overlap. The third freeze--out line then 
has to intersect within the experimental uncertainty and for the same 
value of $\langle B \rangle$. In Fig.~\ref{mutfixratio} global 
values for the triplet of particle ratios have been chosen to demonstrate
the variation of the freeze--out lines with the radius parameter.

\subsection{Centrality Dependence of the Freeze--out Parameters}

The inclusive particle ratios of the existing data base represent averages
over the impact parameter of the collision. 
Prior to the analysis of the inclusive data 
we therefore address the centrality dependence of the
freeze--out parameters.
 
The ratios $\langle M_{\pi^0} \rangle /\langle B \rangle$ and 
$\langle M_{\eta} \rangle/ \langle M_{\pi^0} \rangle$ have been measured 
as a function of the centrality of the collision 
in Ar~+~Ca \cite{MAR97} and Au~+~Au \cite{WOL98}, both at 0.8$A$~GeV.
While in the intermediate--mass Ar~+~Ca system the meson multiplicities 
per participant 
nucleon essentially do not depend on the impact parameter, a clear 
effect is observed in the heavy system Au~+~Au.  
>From peripheral 
($\langle B \rangle$ = 40 $\pm~10$) to central collisions 
($\langle B \rangle$ = 345 $\pm~25$), the neutral--pion multiplicity 
per participant
nucleon $\langle M_{\pi^0} \rangle /\langle B \rangle$ increases
by a factor of about 1.6, while simultaneously the $\eta$--meson multiplicity
per participant nucleon $\langle M_\eta \rangle /\langle B \rangle$
increases by a factor of 3.6 (see Tab.~\ref{aucendeptab}).
The enhancement of the $\eta$--meson production in central collisions
has been attributed to secondary 
reactions involving resonances as intermediate energy storage \cite{WOL98}. 
This interpretation is particularly relevant for the large system size and
the subthreshold incident energy of the measurement and indicates that 
dynamical effects may still be visible in these Au~+~Au collisions.

\begin{figure}
\begin{center}
\mbox{\epsfig{file=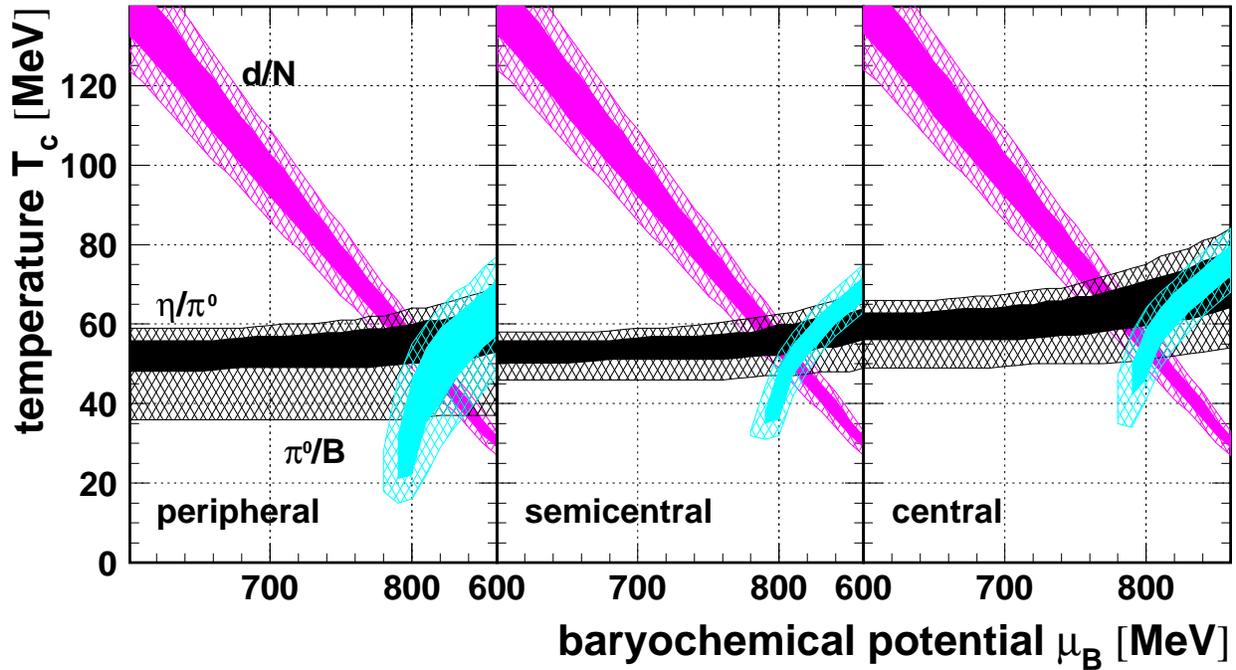,width=\columnwidth}}
\caption[]{Determination of the chemical freeze--out parameters in the 
system Au~+~Au at 0.8$A$~GeV beam energy for peripheral 
($\langle B \rangle$ = 40 $\pm$ 10, left),
semicentral ($\langle B \rangle$ = 227 $\pm$ 20, middle), and central 
collisions ($\langle B \rangle$ = 345 $\pm$ 25, right).
For given impact parameter
the particle ratios $\pi^0 / \rm{B}$, 
$\eta / \pi^0$, and $\rm{d} / \rm{N}$ define three bands, shown in different
greyscales, in the $\mu_B,T_c$--plane. 
The solid (hatched) bands reflect $1\sigma$ ($2\sigma$) intervals
in the experimental uncertainties. 
The freeze--out parameters are determined by the overlap of
the three bands. 
}
\label{aucendepfig}
\end{center}
\end{figure}

Fig.~\ref{aucendepfig} illustrates the thermal analysis of Au~+~Au for 
peripheral, semicentral, and central collisions. Shown are the three bands in 
the $\mu_B$,$T_c$--plane corresponding to the three particle ratios for 
1$\sigma$ and 2$\sigma$ uncertainties. 
It is important to understand the 
correlations between the particle ratios and the parameters of the thermal
model.
One observes that in the vicinity of the common intersection point of the 
bands the ratio $\langle M_{\pi^0} \rangle /\langle B \rangle $ 
is equally sensitive to $\mu_B$ and $T_c$.
Larger values of $\langle M_{\pi^0} \rangle /\langle B \rangle$
mean larger temperatures and smaller chemical potentials of the hadron gas.
This reflects the fact that with increasing $\mu_B$ the baryon density
grows while the density of free pions remains unaffected.
The ratio $\langle M_{\eta} \rangle / \langle M_{\pi^0} \rangle $, on the 
other hand, is practically only 
sensitive to $T_c$. Larger values of
$\langle M_{\eta} \rangle /\langle M_{\pi^0} \rangle $ 
require larger values for $T_c$, while $\mu_B$ is uncritical.
This is due to the fact that producing an $\eta$ meson or exciting an
$N(1535)$ resonance requires more energy and thus a higher temperature
than producing a pion or exciting a $\Delta(1232)$ resonance because
of the pronounced mass difference between both mesons or both resonances,
respectively.
The third ratio,  $\langle M_{\rm{d}} \rangle / \langle M_{\rm{N}} \rangle $,
derives its sensitivity to $\mu_B$ from the difference in baryon number
of the two particles, while the temperature dependence is governed  
by the mass difference between the particles.
With increasing baryochemical potential the temperature of the system
has to drop in order to maintain a given 
$\langle M_{\rm{d}} \rangle / \langle M_{\rm{N}} \rangle $ ratio.
It should be noted, however, that the sensitivity of the freeze--out parameters 
with respect to a variation of 
$\langle M_{\rm{d}} \rangle / \langle M_{\rm{N}} \rangle $ is small.
Modifying the ratio by as much as $\pm$20\% changes 
the freeze--out parameters by not more than 5\%.
Nevertheless, $\langle M_{\rm{d}} \rangle / \langle M_{\rm{N}} \rangle$
provides a useful additional constraint  
in those cases where the neutral--meson data alone would leave room for large
variations in $\mu_B$ and $T_c$.

Fig.~\ref{aucendepfig} shows that in 0.8$A$~GeV Au~+~Au
the experimental particle ratios do define unique sets 
of $\mu_B$ and $T_c$ for each of the three impact parameter selections.
The numerical values as determined by $\chi^2$ minimization
are summarized in Tab.~\ref{aucendeptab}.
The uncertainties quoted represent 
$1\sigma$ standard deviations and reflect the size of the error ellipse 
of the $\mu_B, T_c$--pair at $\chi^2$ = $\chi^2_{min}$ + 1.   
Within these uncertainties the 
baryochemical potential and the chemical freeze--out temperature
obtained in the most peripheral and central collisions, respectively,
agree with the semicentral values of  
$\mu_B = 812 \pm 5$~MeV and $T_c = 52 \pm 2$~MeV.
The ensuing baryon densities $\rho_B$
relative to the nuclear ground--state density $\rho_0$ are also given in 
Tab.~\ref{aucendeptab}.

\begin{figure}
\begin{center}
\mbox{\epsfig{file=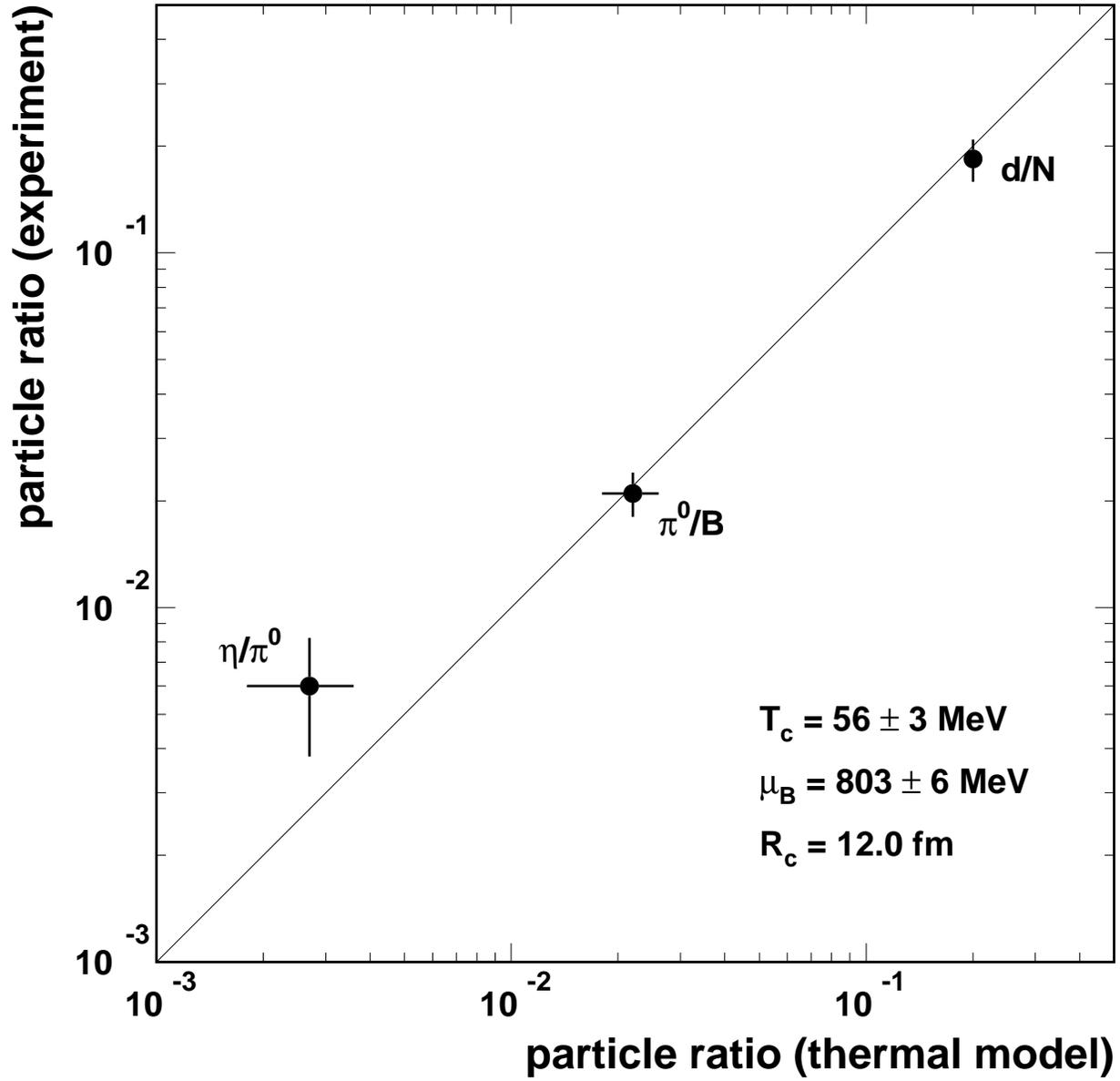,width=\columnwidth}}
\caption[]{
Comparison between experimental particle ratios and ratios as calculated from
the baryochemical potential $\mu_B$ and the temperature $T_c$ at freeze--out. 
The case shown corresponds to central Au~+~Au collisions at 0.8$A$~GeV with 
baryon number $\langle B \rangle$ = 345 $\pm$ 25. $R_c$ is the freeze--out 
radius.
}
\label{expvsmod}
\end{center}
\end{figure}

In our ansatz we are able to describe the central Au~+~Au collisions 
at 0.8$A$~GeV with one common set of freeze--out parameters, see 
Fig.~\ref{expvsmod}, 
while Cleymans et al.~\cite{CLE99} 
emphasize that they cannot reproduce the ratio 
$\langle M_{\eta} \rangle/ \langle M_{\pi^0} \rangle$ within their
hadron--gas model for the same system at 1.0$A$~GeV.
The present analysis is based on the directly measured 
ratio $\langle M_{\eta} \rangle/ \langle M_{\pi^0} \rangle$ for experimentally 
determined values of $\langle B \rangle$, while the authors of 
ref.~\cite{CLE99} have to overcome the difficulty that at 1.0$A$~GeV the ratio 
$\langle M_{\eta} \rangle/ \langle M_{\pi^0} \rangle$ 
has not been measured for truely central collisions. They extrapolate 
$\langle M_{\eta} \rangle/ \langle M_{\pi^0} \rangle$
measured in centrality--biased Au~+~Au collisions at 1.0$A$~GeV beam energy 
\cite{BER94} (c.f. Tab.~\ref{data}) to fully central collisions using 
the centrality dependence of that ratio measured at 0.8$A$~GeV 
\cite{WOL98}. It is known \cite{WOL98,MUE97}, however, that 
the dependence of the meson multiplicity on the number of baryons 
in the fireball is governed by the transverse--mass
excess $\langle m_t \rangle - (\sqrt{s} - 2m_N)$, giving rise to a steeper
increase of the $\eta$ multiplicity with centrality in the more 
subthreshold Au~+~Au collisions at 0.8$A$~GeV. 
With their approach Cleymans et al.~thus overestimate
the $\langle M_{\eta} \rangle/ \langle M_{\pi^0} \rangle$
ratio since at 1.0$A$~GeV the production 
of $\eta$ mesons is considerably less subthreshold than at 0.8$A$~GeV.
In addition, Cleymans et al.~were 
not aware of the fact that the 1.0$A$~GeV measurement is already 
centrality--biased, the value of (1.4 $\pm$ 0.6)\% corresponding
to $\langle B \rangle = 164 \pm 20$ (c.f. Tab.~\ref{data}). Judging from 
the full variation of 40\% observed for the centrality dependence of
$\langle M_{\eta} \rangle/ \langle M_{\pi^0} \rangle$ at 1.0$A$~GeV (see
ref.~\cite{BER94}) we obtain 
$\langle M_{\eta} \rangle/ \langle M_{\pi^0} \rangle$ = (1.8 $\pm$ 0.7)\% 
for central collisions. Thus, the lower boundary of the 2$\sigma$ 
uncertainty band for a revised $\eta/\pi^0$ freeze--out line may reach 
$T_c$ = 55 MeV at $\mu_B$ = 800 MeV (see Figs.~\ref{mutfixradius} 
and \ref{mutfixratio}), well within the 2$\sigma$ ellipsoid of
the intersecting remaining freeze--out lines in the analysis of
Cleymans et al.~\cite{CLE99}.

In summary, no significant centrality dependence of the freeze--out parameters
is found for the heavy system Au~+~Au at 0.8$A$~GeV. 
For the lighter systems and for beam energies which are less subthreshold
for $\eta$ production a possible dependence on the impact parameter
therefore can be neglected at the present level of accuracy.

\subsection{Results for Inclusive Collisions}

\begin{figure}
\begin{center}
\mbox{\epsfig{file=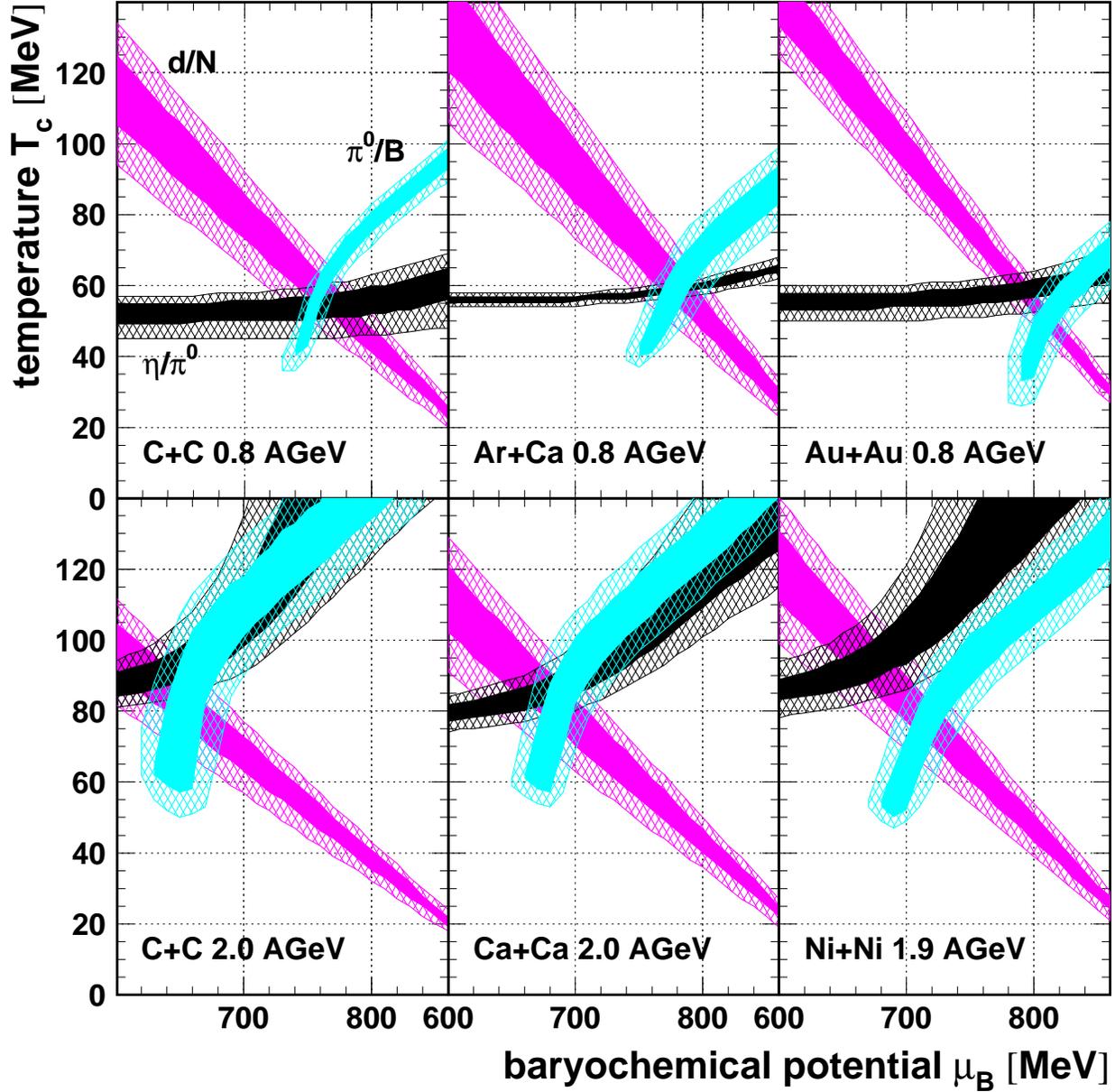,width=\columnwidth}}
\caption[]{Determination of the chemical freeze--out 
parameters T$_c$ and $\mu_B$. 
For given
system and beam energy the particle ratios $\pi^0 / \rm{B}$, 
$\eta / \pi^0$, and $\rm{d} / \rm{N}$ define three bands, distinguished by
different greyscales, in the $\mu_B,T_c$--plane. 
The solid (hatched) bands reflect $1\sigma$ ($2\sigma$) intervals
in the experimental uncertainties. 
The freeze--out parameters are determined by the overlap of
the three bands. The individual panels show the solutions
obtained for the lightest, an intermediate--mass, and the heaviest system 
studied at the lowest and highest beam energies, respectively.}
\label{mut}
\end{center}
\end{figure}

Fig.~\ref{mut} shows the results of the thermal analysis for the lightest, an
intermediate--mass, and the heaviest system, both at the lowest and highest 
beam energy studied.
Presented in the $\mu_B$,$T_c$--plane are the bands defined by the input values
$\langle M_{\pi^0} \rangle /\langle B \rangle$, 
$\langle M_{\eta} \rangle/ \langle M_{\pi^0} \rangle$, and
$\langle M_{\rm{d}} \rangle / \langle M_{\rm{N}} \rangle$.
In all cases the three bands overlap and, therefore, define the
freeze--out parameters $\mu_B$ and $T_c$ in an unambiguous way.
Even collision systems with as few participating nucleons
as C~+~C seem to comply with the model--assumption of 
chemical equilibrium.
The results for the full set of inclusive measurements are summarized in 
Tab.~\ref{resmut}.
The analysis reveals a systematic 
reduction of the baryochemical potential from 800 to 650~MeV 
with increasing beam energy, which is accompanied by an 
increase of the freeze--out temperature from 55 to 90~MeV. For given beam 
energy the chemical potential $\mu_B$ grows with increasing system 
size while the freeze--out temperature $T_c$ stays almost constant.
These findings agree with 
results quoted from similar analyses \cite{HON98,CLE98,CLE99}, although we 
observe a general trend towards smaller baryochemical potentials and 
higher temperatures as compared to \cite{CLE98,CLE99}. 

The freeze--out radius $R_c$ is fixed by the average
number of baryons through Eq.~\ref{radiusfix} and 
enters the model analysis via the surface correction term. 
$R_c$ is larger for the heavy target--projectile 
combinations, but increases less than the trivial 
$\langle B \rangle^{1/3}$ law, and it decreases with increasing beam energy.
The corresponding baryon densities at chemical freeze--out are in
general smaller than about half the nuclear ground--state density which 
is in good agreement with results from other analyses  
\cite{BRA95,BRA96,HON98}. 

\subsection{Chemical Composition of the Fireball}

\begin{figure}
\begin{center}
\mbox{\epsfig{file=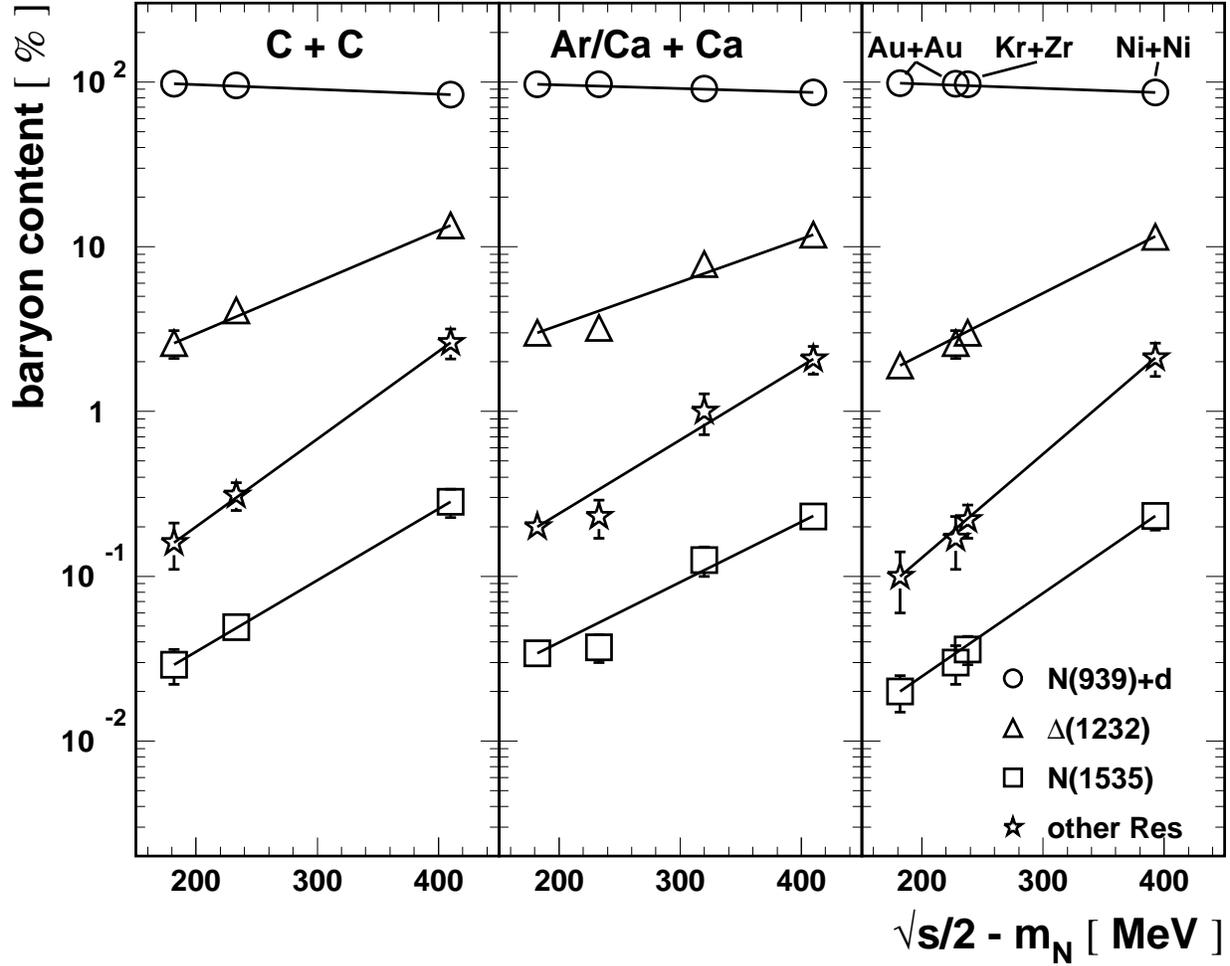,width=\columnwidth}}
\caption[]{Baryon content at chemical freeze--out in the hadron--gas
model. The relative contributions to the total baryon number 
from nucleons and deuterons,
$\Delta(1232)$ resonances, $N(1535)$ resonances, and the sum of all 
remaining $\Delta$ and $N$ resonances are plotted as a function of the 
energy available per baryon. The three panels show from
left to right the results obtained for light, intermediate--mass, and heavy 
systems. 
The error bars reflect the uncertainty in the 
freeze--out parameters $\mu_B$ and $T_c$ (see Tab.~\ref{resmut}). 
Straight lines are drawn to guide the eye.}
\label{comp}
\end{center}
\end{figure}

The quantities $\mu_B$ and $T_c$ are the two important free parameters 
of our model. 
They essentially determine the properties of the fireball. As an example, 
Fig.~\ref{comp} shows the chemical composition of the baryon sector at 
chemical freeze--out as a function 
of the energy available in the nucleon--nucleon system. The system 
size is a further parameter. Light, intermediate--mass, 
and heavy systems are therefore treated separately in Fig.~\ref{comp}.

Immediately apparent is the pronounced energy dependence 
of the chemical composition, while for any given incident energy
the chemical composition is nearly independent on the size of the system. 
The fraction of baryons excited to resonance states grows from 2--3\%
at 0.8$A$~GeV beam energy to about 15\% around 2.0$A$~GeV.
The $\Delta(1232)$ resonance is populated most abundantly.
The ratio of heavier resonances to the $\Delta(1232)$ resonance, 
however, increases from about 7\% at 0.8$A$~GeV to 
nearly 20\% around 2.0$A$~GeV, indicative of a more equal population of the 
resonance spectrum at higher incident energies. 
As illustrated by the straight lines in Fig.~\ref{comp}, the energy dependence 
of the individual relative baryon populations is in reasonable agreement 
with exponential behavior.
The slopes of these exponentials are nearly equal for $\Delta(1232)$ and 
$N(1535)$ and steeper for the summed contributions of the
remaining $\Delta$ and $N$ resonances.

The baryon composition of the chemical freeze--out state being established 
it is interesting to extrapolate back to the high--density phase of 
the collision. Around 2$A$~GeV beam energy 
microscopic model calculations quite consistently give a maximum baryon 
density of $\rho_{max}$~$\approx$~2.5~$\rho_0$ and predict a ratio of
0.3 to 0.4 for the ratio of the number of $\Delta(1232)$ resonances at 
freeze--out to the corresponding number at maximum 
compression \cite{TEI97,WEB96}. Using this information together with an 
allowance for contributions from higher resonances, one obtains a value 
of $0.4~\rho_{max} \approx \rho_0$
for the maximum resonance density in the collisions at 1.9$A$ and 2.0$A$~GeV,
see Fig.~\ref{comp}. Although the density of baryon resonances only amounts 
to about 40\% of the
total baryon density, resonance--resonance interactions might take place.
Hadronic matter in that state has been referred to as resonance matter 
\cite{EHE93,MOS93}.

\begin{figure}
\begin{center}
\mbox{\epsfig{file=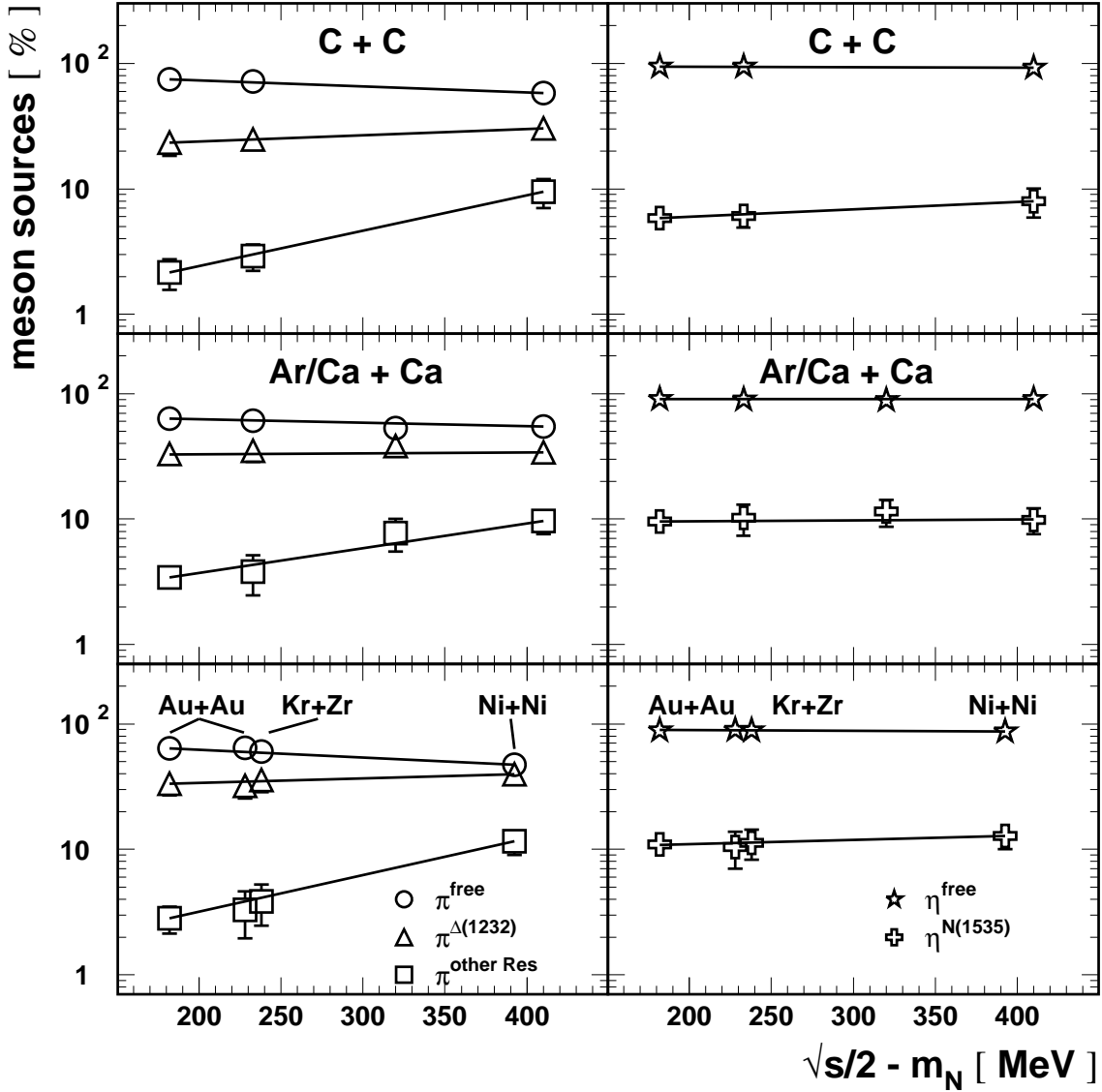,width=\columnwidth}}
\caption[]{Pion (left) and $\eta$ meson sources (right) at chemical 
freeze--out in the hadron--gas model. The mesons are either present as
free particles or still bound in resonance states. The corresponding relative 
contributions are shown as a function of the energy available per baryon 
for light (top), intermediate--mass (middle), and heavy systems (bottom). 
The error bars reflect the uncertainty in the 
freeze--out parameters $\mu_B$ and $T_c$ (see Tab.~\ref{resmut}). 
Straight lines are drawn to guide the eye.}
\label{mesonsrc}
\end{center}
\end{figure}

Concerning the meson sector of the freeze--out state, an important 
observation is that a sizeable 
fraction of the mesons are present as free mesons in chemical equilibrium 
with the baryons. This is immediately apparent, if one compares the 
population of $\Delta(1232)$ and $N(1535)$
resonances with the $\pi^0$ and $\eta$ multiplicities observed
asymptotically (see Tab.~\ref{data}). For a quantitative discussion, we 
plot in Fig.~\ref{mesonsrc} the
fraction of free mesons and the fraction of mesons bound in resonances
at freeze--out as a function of the energy available in the nucleon--nucleon
system, again differentiating between light, intermediate--mass, and 
heavy systems.
The relative contributions to the final--state pion yields reveal a moderate
energy dependence,
the fraction of free pions decreasing from about 65\% at 0.8$A$~GeV
beam energy to about 55\% at 2.0$A$~GeV. Pions from 
the $\Delta(1232)$ resonance behave in a complementary way, as expected, with 
contributions of about 30\% at 0.8$A$~GeV and about 40\% at 2.0$A$~GeV.
The remaining intensity can be attributed to heavier resonances which,
according to their higher mass, give rise to a steeper energy dependence 
than exhibited by pions from $\Delta(1232)$--resonance decays.
For $\eta$ mesons resonance decay after chemical freeze--out is unimportant.
About 85--90\% of the asymptotically observed mesons are already free
at chemical freeze--out. 
The energy dependence of the $N(1535)$ resonance contribution is comparable 
to that of the pion contribution from $\Delta(1232)$ decays (see also
Fig.~\ref{comp}).
For pions as well as for $\eta$ mesons 
a trend towards larger contributions from resonance decays  
is visible in the heavier systems.

\subsection{Time Order of Chemical and Thermal Freeze--out}

In a sudden freeze--out scenario the temperature of an equilibrated hadron gas  
at thermal freeze--out is characterized by the momentum
spectra of the emitted particles.
The experimental spectra, however, may be modified 
by resonance decays and, in particular in heavy systems, by the rescattering
of particles
off spectator material. In addition, energy transferred into collective flow 
reduces the temperature of the system. Thus, in order 
to extract the freeze--out temperature $T_{th}$ from the spectral 
shapes, further assumptions have to be made leading to 
model--dependent results.
Nevertheless, the meson spectra of the present data base do provide 
a valuable consistency check for the chemical freeze--out analysis in the sense 
that the different model temperatures are subject to constraints
due to the time order of chemical and thermal freeze--out.

\begin{figure}
\begin{center}
\mbox{\epsfig{file=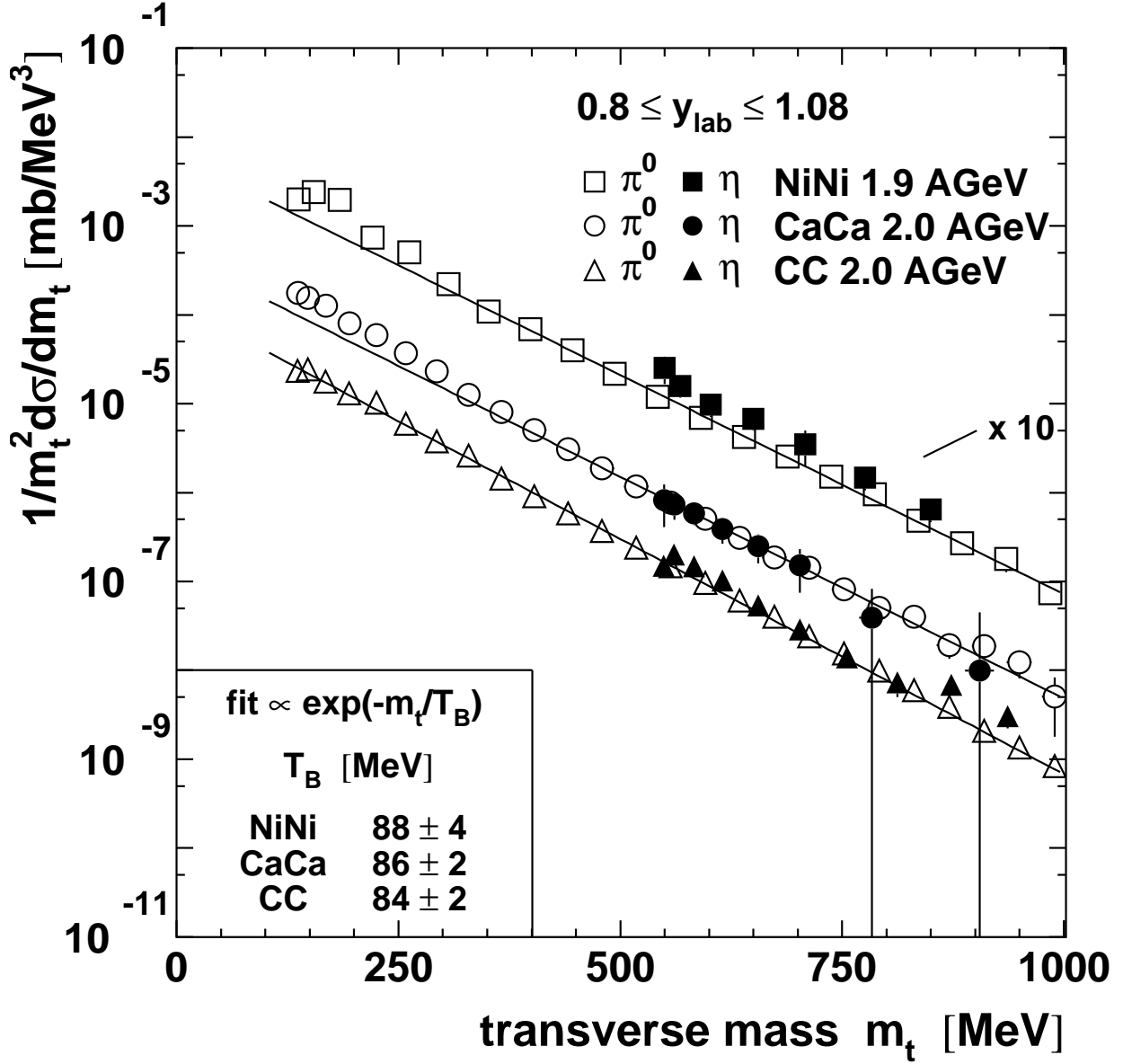,width=\columnwidth}}
\caption[]{Impact--parameter inclusive 
transverse--mass spectra of $\pi^0$ and $\eta$--mesons as observed
in the systems C~+~C \cite{AVE97} and Ca~+~Ca \cite{VOG98} at 2.0$A$~GeV 
beam energy and in Ni~+~Ni at 1.9$A$~GeV \cite{APP97}. The distributions 
are divided by the square of the transverse mass. In this representation 
midrapidity particles from a thermal source are expected to exhibit a purely 
exponential spectrum. The solid lines represent Boltzmann fits 
(see Eq.~\ref{mtdis})
to the $\pi^0$ 
data for $m_t \ge 400$~MeV.}
\label{mtscal}
\end{center}
\end{figure}

As an example we consider the impact--parameter inclusive transverse--mass 
spectra of $\pi^0$ and $\eta$ mesons presented in Fig.~\ref{mtscal}.
The spectra can be described reasonably well by exponential 
distributions, provided one excludes the low transverse
masses from the fit.
For all three collision systems 
the Boltzmann parameters $T_B$, which reflect the temperature 
$T_{th}$ of the fireball at thermal freeze--out in case of vanishing 
collective flow, agree with the 
corresponding chemical freeze--out temperatures $T_c$. 
A similar quality of agreement is observed for Ar~+~Ca at 1.5$A$~GeV, 
see Tab.~\ref{resmut}.
At the lower beam energies of 1.0$A$ and 0.8$A$~GeV the experimental
situation for $T_B$ seems unclear at present, except for the heavy systems 
Kr~+~Zr and Au~+~Au. In Au~+~Au at 0.8$A$~GeV we have $T_B$ $>$ $T_c$ 
at the 1$\sigma$ level, while at 1.0$A$~GeV the Boltzmann temperatures in both 
systems are significantly higher than the corresponding chemical freeze--out
temperatures, see Tab.~\ref{resmut}.

The absence of a low--m$_t$ enhancement in the C~+~C spectrum of 
Fig.~\ref{mtscal} is remarkable. To the extent that rescattering can be 
assumed negligible, the light C~+~C system should show the expected
influence of resonance decay. Instead,  
a perfectly exponential behavior is observed over the full m$_t$ range,
although pions from resonance decay do comprise 40\% of the 
total pion intensity (see Fig.~\ref{mesonsrc}). Consequently,
the onset of the low--m$_t$ enhancement visible in the Ca~+~Ca and Ni~+~Ni 
cases of Fig.~\ref{mtscal} probably
has to be attributed to rescattering in spectator material.

\begin{figure}
\begin{center}
\mbox{\epsfig{file=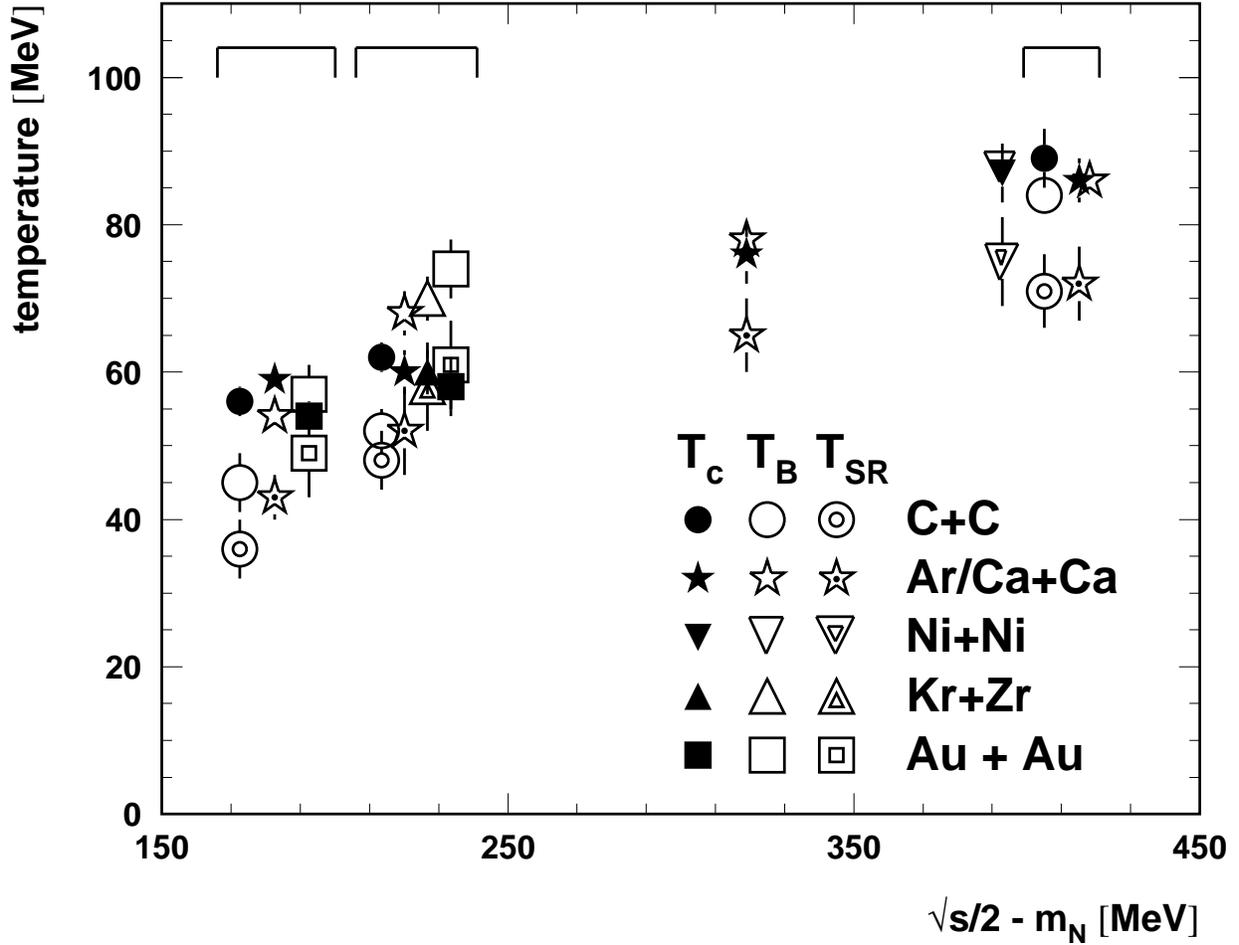,width=\columnwidth}}
\caption[]{
Chemical ($T_c$) and model--dependent thermal ($T_B$ and $T_{SR}$) freeze--out
temperatures as a function of the energy available per baryon.
The data points at 0.8$A$~GeV, 1.0$A$~GeV, and 2.0$A$~GeV beam energy
are slightly shifted in energy with respect to their nominal position
(center of brackets).}
\label{temperatures}
\end{center}
\end{figure}

Collective flow affects the meson spectra. The point 
is, that one has to know the underlying flow profile
in order to analyze a given spectral shape. For the present analysis 
we have chosen the blast model proposed by Siemens and Rasmussen
\cite{SIE79}. In this model, the fireball in thermal 
equilibrium is assumed to expand isotropically.
All particles in the fireball share a common temperature $T_{SR}$ 
and have a common radial--flow velocity $\beta_{SR}$.
The modification of the spectra compared to pure Boltzmann distributions
becomes more significant the heavier the considered particle species is.
Since pions and $\eta$ mesons have low masses their spectra are not very 
sensitive to the radial--flow velocity. Therefore, a fit to the 
mesonic transverse--mass spectra considering both $T_{SR}$ and $\beta_{SR}$
as free parameters would determine the freeze--out parameters
only with large uncertainties.
To avoid this situation, we exploit the fact that for the present range of
collision systems $\beta_{SR}$ is known to have values  
between 0.25 and 0.35 in accord with the systematics of 
radial flow velocities measured for heavier particles 
(A=1 to 4) in central collisions as quoted in \cite{HER96}.
Thus only $T_{SR}$ is treated as a fit parameter while $\beta_{SR}$ 
is kept constant. 
In the $\pi^0$ cases we furthermore restrict the fit to
$m_t \ge 400$~MeV as was done for the Boltzmann fits.
Under these conditions the $\pi^0$ and $\eta$ spectra can be described
within the Siemens--Rasmussen model for all beam energies 
and target--projectile combinations.
In general, the extracted temperatures $T_{SR}$ are 10--20\% smaller
than the corresponding Boltzmann temperatures 
$T_B$, see Fig.~\ref{temperatures}.   
The actual reduction is correlated with the magnitude of 
$\beta_{SR}$ chosen in the fit, with $T_{SR}$ decreasing for 
increasing $\beta_{SR}$.

Thermal freeze--out does not occur before chemical 
freeze--out and, therefore, the thermal 
freeze--out temperature $T_{th}$ cannot be larger than $T_c$.
The available midrapidity spectra of 
the $\pi^0$ and $\eta$ mesons do support this conjecture. If one
neglects radial flow, the spectral shapes observed in the light and 
intermediate--mass systems provide inverse--slope parameters
which are in accord with the chemical freeze--out temperatures $T_c$
derived from the particle yields. 
Taking radial flow into account results in slightly lower 
temperatures at thermal freeze--out compared to chemical freeze--out,
indicative of cooling as the system develops in time.
The meson spectra observed in the heavy systems Kr~+~Zr and Au~+~Au, 
in particular, do require the inclusion of  
flow in order to achieve consistency ($T_{c}$ $\geq$ $T_{th}$) between thermal 
and chemical freeze--out temperatures.

\subsection{Chemical Freeze--out Curve for Hadronic Matter}

\begin{figure}
\begin{center}
\mbox{\epsfig{file=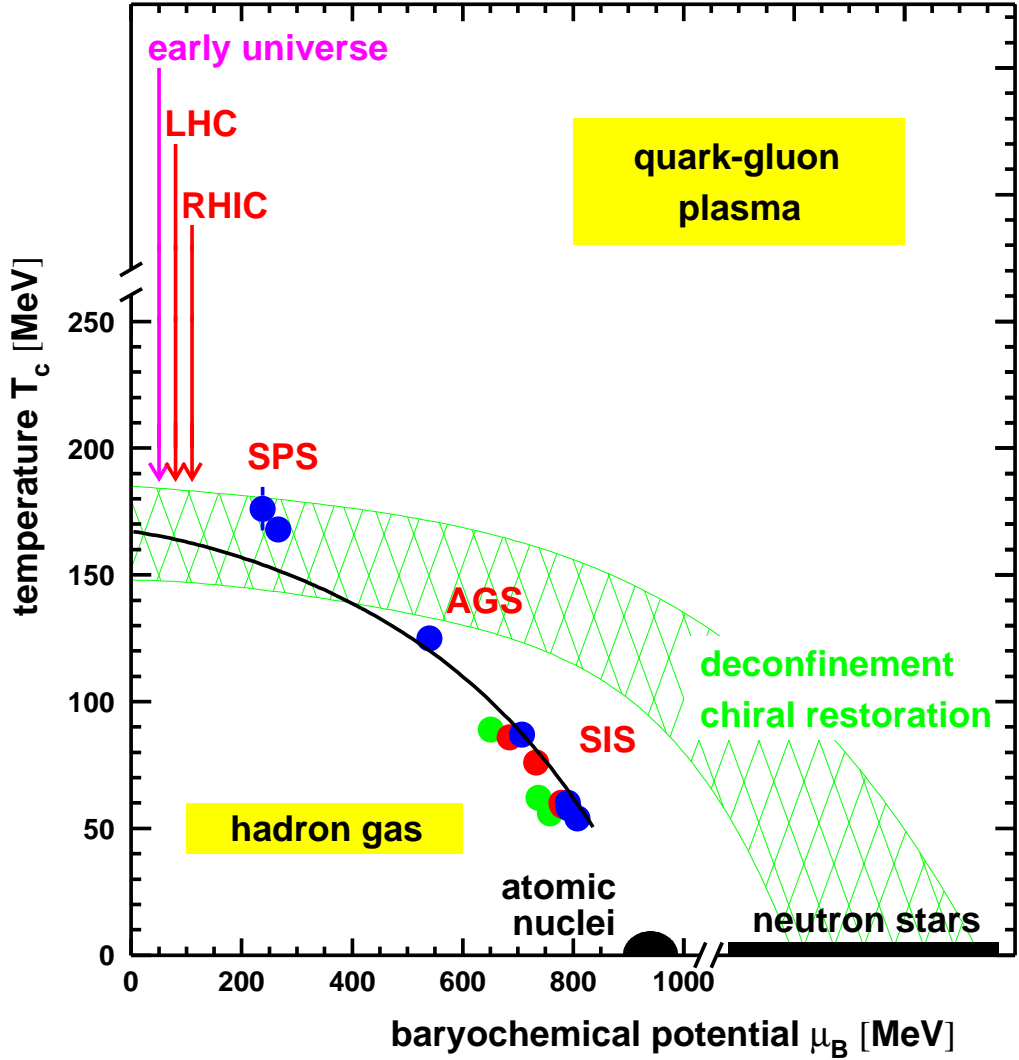,width=145mm}}
\caption[]{Phase diagram of hadronic matter. The chemical freeze--out
temperatures $T_c$ are shown as a function of the corresponding
baryochemical potentials $\mu_B$ as obtained from yield ratios of particles
produced in nucleus--nucleus collisions at various incident energies. Results 
of the present work are plotted together with AGS and SPS results 
\cite{BRA95,BRA96}. The grayscale of the SIS data points reflects 
the system size, i.e.~light (dark) points correspond to light (heavy) 
systems. The solid curve (taken from Ref.~\cite{CLR99})
approximates the curve 
for chemical freeze--out of hadronic matter. It corresponds to an average
energy per hadron of 1~GeV in a hadron--gas model, 
while the crosshatched band indicates location and uncertainty of the 
boundary between a gas of hadrons and a plasma of gluons and deconfined 
quarks.}
\label{phase}
\end{center}
\end{figure}

Fig.~\ref{phase} shows our results within a schematic phase diagram of
hadronic matter, together with other data points obtained from
particle--production experiments at higher energies \cite{BRA95,BRA96}.
In contrast to the AGS and SPS results the chemical freeze--out
parameters deduced at SIS energies are far below the expected phase boundary 
between hadron gas and quark--gluon plasma.
For zero chemical potential the critical temperature is constrained by 
lattice QCD calculations. 
For finite values of $\mu_B$ the phase boundary can be approximated in 
a simple model by equating chemical potential and pressure in the
hadronic phase and in an idealized quark--gluon plasma \cite{BRA96}.
It has recently been noticed by Cleymans and Redlich that, within their
hadron--gas model, the chemical freeze--out curve corresponds to an 
average energy per hadron of 1~GeV \cite{CLR99}.
The resulting curve is shown in Fig.~\ref{phase} as a solid line which
is in close agreement with the deduced freeze--out parameters.

At SIS energies thermal and chemical 
freeze--out seem to nearly coincide. 
In contrast, at SPS beam  
energies the freeze--out parameters clearly indicate 
that thermal freeze--out reflects a later stage of the collision 
as significantly lower temperatures compared to chemical 
freeze--out are deduced \cite{STO98}.

\section{Summary}

The inclusive neutral--pion and $\eta$--meson yields measured in 
symmetric collisions of
light, intermediate--mass, and heavy nuclei near midrapidity in the
energy range from 0.8$A$~GeV to 2.0$A$~GeV are consistent with the
formation of a hadronic fireball in chemical equilibrium,
as described by an isospin--symmetric ideal hadron gas.
With increasing bombarding energy the
baryochemical potential $\mu_B$ decreases from 800~MeV to 650~MeV 
while simultaneously the temperature $T_c$ increases from 55~MeV to 90~MeV.
Concerning the system--size dependence, we find that $\mu_B$ grows with 
increasing mass of the projectile--target combination while $T_c$ remains 
about constant. The centrality dependence of $\mu_B$ and $T_c$ 
has been investigated in the system Au~+~Au at 0.8$A$~GeV. Here 
the $\eta/\pi^0$ ratio is expected to be most susceptible to the 
impact parameter, both 
because of the large mass of the system and because of the low incident 
beam energy. No centrality dependence is observed.

Apart from the moderate system size dependence, the freeze--out 
parameters $\mu_B$ and $T_c$ completely characterize the
system. Given $\mu_B$ and $T_c$ one can calculate the 
hadrochemical composition of the fireball.
While at 0.8$A$~GeV beam energy only 2--3\% of the nucleons are
excited to resonance states at chemical freeze--out, this fraction 
increases to about 15\% at 2.0$A$~GeV bombarding energy.
According to transport--model calculations, this resonance content at 
freeze--out implies a resonance density in the  
high--density phase of the collision of about normal nuclear matter 
density, thus justifying the term resonance matter for the specific
state of hadronic matter created in heavy--ion collisions at SIS energies.
In the meson sector of our model about 50\% of all 
asymptotically observed pions are still bound in resonance states 
at chemical freeze--out, while only about 10\% of the 
final--state $\eta$ mesons are bound at that moment. 

In contrast to the particle yields, which convey the status 
at chemical freeze--out, the shapes of the related transverse--mass spectra
do reflect thermal freeze--out. After proper allowance for 
radial flow the slope parameters observed for the midrapidity spectra of the 
neutral mesons correspond to thermal freeze--out temperatures which 
are equal to or slightly lower than the chemical freeze--out temperatures. 
In contrast to ultrarelativistic collisions studied at the 
SPS, chemical and thermal freeze--out thus seem to occur almost 
simultaneously at SIS energies as is also 
observed for heavy--ion collisions at the AGS.

Particle yields and spectral shapes reflect only two facets of the very 
complex process of a relativistic nucleus--nucleus collision.
With these observables alone it is not possible to decide whether 
chemical equilibrium is actually reached 
during any stage of such a collision. In fact, the observation 
that for the heavy collision systems the agreement between data and 
hadron--gas model is not as good as for 
the light systems can be seen as one indication that physics 
beyond equilibrium concepts may still be visible in the final 
state of relativistic nucleus--nucleus collisions around 1$A$~GeV.

\acknowledgements

The extensive data base on neutral--meson production in heavy--ion collisions 
as accumulated by the TAPS collaboration has been instrumental for this study.
It is a pleasure to acknowledge helpful discussions with
many colleagues, in particular with P.~Braun--Munzinger, J.~Cleymans,
H.~Oeschler, K.~Redlich, and P.~Senger. 
One of us (R.A.) is supported by the A.~v.~Humboldt Foundation as 
an F.~Lynen fellow.

\begin{table}
\caption[]{Mean inclusive $\pi^0$ multiplicities $\langle M_{\pi^0} \rangle $ 
relative to the average number of participating nucleons 
$\langle B \rangle $ 
and mean inclusive $\eta$ multiplicities $\langle 
M_{\eta} \rangle $ relative to $\langle M_{\pi^0} \rangle $ measured for 
various nucleus--nucleus collisions in the beam--energy range
from 0.8$A$~GeV to 2.0$A$~GeV. The last column denotes interpolated
mean deuteron multiplicities $\langle M_{\rm{d}} \rangle $ relative to the 
mean nucleon multiplicity $\langle M_{\rm{N}} \rangle $.
The data for Kr~+~Zr and Au~+~Au (both at 1$A$~GeV, see \cite{SCH94}) have 
been revised to account for the fact that the first--level trigger condition
in effect in these measurements introduced a bias towards centrality. 
}
\label{data}
\begin{tabular}{d r@{~+~}l 
c
r@{}l@{${}\pm{}$}r@{}l r@{}l@{${}\pm{}$}r@{}l
r@{}l@{${}\pm{}$}r@{}l}
\multicolumn{1}{c}{$E_{beam}$ [$A$~GeV]} & 
\multicolumn{2}{c}{System} & 
$\langle B \rangle$        & 
\multicolumn{4}{c}{$\frac{\langle M_{\pi^0} \rangle }
{\langle B \rangle }$ [\%]} & 
\multicolumn{4}{c}{$\frac{\langle M_{\eta} \rangle }{\langle M_{\pi^0} 
\rangle }$ [\%]} & 
\multicolumn{4}{c}{$\frac{\langle M_{\rm{d}} \rangle }{\langle M_{\rm{N}} 
\rangle }$ [\%]} \\
\tableline
0.8 & C  & C  &  6          & 3&.7 & 0&.3 & 0&.31 & 0&.11 & 11&.3 & 2&.7 \\
0.8 & Ar & Ca &  20         & 3&.1 & 0&.5 & 0&.41 & 0&.04 & 16&.3 & 3&.9 \\
0.8 & Au & Au &  125$\pm$15 & 1&.6 & 0&.3 & 0&.38 & 0&.08 & 18&.4 & 2&.5 
\\[1.8ex]
1.0 & C  & C  &  6          & 5&.6 & 0&.4 & 0&.57 & 0&.14 & 10&.6 & 2&.5 \\
1.0 & Ar & Ca &  20         & 3&.0 & 0&.3 & 1&.3  & 0&.8  & 15&.3 & 3&.7 \\
1.0 & Kr & Zr &  79$\pm$9   & 2&.5 & 0&.7 & 1&.3  & 0&.6  & 17&.7 & 2&.3 \\
1.0 & Au & Au &  164$\pm$20 & 2&.3 & 0&.5 & 1&.4  & 0&.6  & 17&.2 & 2&.3 
\\[1.8ex]
1.5 & Ar & Ca &  20         & 6&.5 & 0&.5 & 2&.2  & 0&.4  & 12&.7 & 3&.0 
\\[1.8ex]
2.0 & C  & C  &  6          &13&.8 & 1&.4 & 3&.6  & 0&.4  &  7&.0 & 1&.7 \\
2.0 & Ca & Ca & 20          &11&.1 & 1&.1 & 2&.7  & 0&.3  & 10&.2 & 2&.5 \\
1.9 & Ni & Ni &  29         & 8&.6 & 0&.9 & 3&.3  & 0&.4  & 12&.7 & 2&.8 \\
\end{tabular}
\end{table}

\begin{table}
\caption[]{
Baryon resonances considered as constituents of the hadronic fireball.
Given for each resonance are the nominal mass m$_{R}$, the total width
$\Gamma_{R}$ at m$_{R}$ as well as the partial widths 
corresponding to the three decay modes considered in the present model.
Values are taken from \cite{PDGBARYON} (3-- or 4--star status) with the 
exception of $\Delta(1232)$ \cite{KOC84} and $N(1535)$ \cite{KRU95}.
}
\label{resonances}
\begin{tabular}{cccrrr}
resonance
&   mass m$_{R}$ [MeV]
&   width $\Gamma_{R}$ [MeV]
& $\Gamma_{1\pi}$/$\Gamma_{R}$ [\%] 
& $\Gamma_{2\pi}$/$\Gamma_{R}$ [\%]
& $\Gamma_{\eta}$/$\Gamma_{R}$ [\%] \\
\tableline
$\Delta$(1232)   &      1232      &      110      & 100      &  0     &  0\\ 
     $N(1440)$   &      1440      &      350      &  65      & 35     &  0\\ 
     $N(1520)$   &      1520      &      120      &  55      & 45     &  0\\ 
     $N(1535)$   &      1544      &      203      &  50      &  0     & 50\\ 
$\Delta$(1600)   &      1600      &      350      &  15      & 85     &  0\\ 
$\Delta$(1620)   &      1620      &      150      &  30      & 70     &  0\\ 
     $N(1650)$   &      1650      &      150      &  80      & 20     &  0\\ 
     $N(1675)$   &      1675      &      150      &  45      & 55     &  0\\ 
     $N(1680)$   &      1680      &      130      &  65      & 35     &  0\\ 
     $N(1700)$   &      1700      &      100      &  10      & 90     &  0\\ 
$\Delta$(1700)   &      1700      &      300      &  15      & 85     &  0\\ 
     $N(1710)$   &      1710      &      100      &  15      & 85     &  0\\ 
     $N(1720)$   &      1720      &      150      &  20      & 80     &  0\\ 
\end{tabular}
\end{table}

\begin{table}
\caption[]{Particle ratios and chemical freeze--out parameters for 
peripheral, semicentral, and central collisions 
in the system Au~+~Au at 0.8$A$~GeV. The impact parameter 
was selected according to the number of 
participating nucleons $\langle B \rangle $ given in the first row.
The next entries denote the meson ratios  
($\langle M_{\pi^0} \rangle / \langle B \rangle$,
$\langle M_{\eta} \rangle / \langle M_{\pi^0} \rangle$), the 
inclusive deuteron/nucleon ratio  
$\langle M_{\rm{d}} \rangle / \langle M_{\rm{N}} \rangle $,
and the freeze--out radius $R_{c}$. 
The subsequent rows show the freeze--out parameters $\mu_B$ 
and $T_c$ together with the resulting baryon density $\rho_B$ 
relative to the nuclear ground--state density $\rho_0$ = 0.168~fm$^{-3}$.
The values are determined by $\chi^2$ minimization, the uncertainties 
represent
$1\sigma$ standard deviations as evaluated from the size of the error ellipse 
of the $\mu_B, T_c$ pair at $\chi^2$ = $\chi^2_{min}$ + 1.   
}
\label{aucendeptab}
\begin{tabular}
{
c                                   | 
r@{$\,\pm\,$}l 
r@{$\,\pm\,$}l 
r@{$\,\pm\,$}l
}
& 
\multicolumn{2}{c}{peripheral}      & 
\multicolumn{2}{c}{semicentral}     & 
\multicolumn{2}{c}{central}                       \\
\tableline
$\langle B \rangle$                 & 
40     &   10                       & 
227    &   20                       & 
345    &   25                                     \\ 
$\langle M_{\pi^0}  \rangle / \langle B          \rangle$ [\%] &
1.3    &   0.4                      & 
1.6    &   0.2                      & 
2.1    &   0.3                                    \\ 
$\langle M_{\eta}   \rangle / \langle M_{\pi^0}  \rangle$ [\%] &
0.27   &   0.13                     & 
0.29   &   0.10                     & 
0.60   &   0.22                                   \\ 
$\langle M_{\rm{d}} \rangle / \langle M_{\rm{N}} \rangle$ [\%] &
\multicolumn{6}{c}{ 18.4 $\pm$ 2.5~~(inclusive)}  \\
$R_c$ [fm ]                         &  
\multicolumn{2}{c}{6.5}             &                
\multicolumn{2}{c}{11.0}            &                
\multicolumn{2}{c}{12.0}                          \\
$\mu_B$ [MeV]                       &
812    &   5                        & 
812    &   5                        & 
803    &   6                                      \\ 
$T_c$ [MeV]                         &  
51     &   2                        & 
52     &   2                        & 
56     &   3                                      \\ 
$\rho_B/\rho_0$                     &
0.21   & 0.07                       & 
0.23   & 0.08                       & 
0.28   & 0.09                                     \\ 
$\chi^2_{min}$                      &
\multicolumn{2}{c}{0.88}            &                
\multicolumn{2}{c}{1.50}            &                
\multicolumn{2}{c}{2.93}                          \\
\end{tabular}
\end{table}

\begin{table}
\caption[]{Chemical freeze--out parameters $\mu_B$ and $T_c$ for
the various systems investigated, together with
the resulting baryon densities $\rho_B$.
The values are determined by $\chi^2$ minimization, uncertainties represent
$1\sigma$ standard deviations as evaluated from the size of the error ellipse 
of the $\mu_B, T_c$ pair at $\chi^2$ = $\chi^2_{min}$ + 1. Also given are the 
model--dependent thermal freeze--out temperatures $T_{B}$ and 
$T_{SR}$ as obtained from Boltzmann fits and 
within the blast model of Siemens and Rasmussen, respectively.
} 
\label{resmut}
\begin{tabular}{
d 
r@{~+~}l|| 
c
r@{$\,\pm\,$}l
r@{$\,\pm\,$}l
r@{$\,\pm\,$}l 
c|
r@{$\,\pm\,$}l
r@{$\,\pm\,$}l
}
\multicolumn{3}{c||}{}                   &
\multicolumn{8}{l|}{chemical freeze--out}&
\multicolumn{4}{l}{thermal freeze--out}  \\
\multicolumn{1}{c}{$E_{beam}$}           & 
\multicolumn{2}{c||}{System}             & 
~$R_c$~                                  & 
\multicolumn{2}{c}{$\mu_B$}              & 
\multicolumn{2}{c}{$T_c$}                & 
\multicolumn{2}{c}{$\rho_B/\rho_0$}      &
~$\chi^2_{min}$~                         &
\multicolumn{2}{c}{$T_{B}$}              &
\multicolumn{2}{c}{$T_{SR}$}             \\ 
\multicolumn{1}{c}{[$A$~GeV]}            & 
\multicolumn{2}{c||}{}                   & 
[fm]                                     & 
\multicolumn{2}{c}{[MeV]}                & 
\multicolumn{2}{c}{[MeV]}                & 
\multicolumn{2}{c}{}                     &
{}                                       &
\multicolumn{2}{c}{[MeV]}                &
\multicolumn{2}{c}{[MeV]}               \\ 
\tableline
0.8    & C  & C                          & 
4.5                                      &    
758    & 5                               & 
56     & 2                               & 
0.09   & 0.03                            & 
0.36                                     &
45     & 4                               &
36     & 4                               \\
0.8    & Ar & Ca                         & 
5.5                                      &    
780    & 7                               & 
59     & 1                               & 
0.19   & 0.05                            & 
0.10                                     &
54     & 2                               &
43     & 3                               \\ 
0.8    & Au & Au                         & 
8.5                                      &    
808    & 5                               & 
54     & 2                               & 
0.27   & 0.08                            & 
5.90                                     &
57     & 4                               &
49     & 6                               \\[1.8ex] 
1.0    & C  & C                          & 
4.5                                      &    
737    & 5                               & 
62     & 2                               & 
0.10   & 0.03                            & 
1.05                                     &
52     & 3                               &
48     & 4                               \\ 
1.0    & Ar & Ca                         & 
5.0                                      &    
779    & 7                               & 
60     & 3                               & 
0.21   & 0.08                            & 
1.14                                     &
68     & 3                               &
52     & 6                               \\ 
1.0    & Kr & Zr                         & 
7.5                                      &    
790    & 7                               & 
60     & 3                               & 
0.27   & 0.09                            & 
2.30                                     &
70     & 3                               &
58     & 6                               \\ 
1.0    & Au & Au                         & 
9.5                                      &    
792    & 7                               & 
58     & 4                               & 
0.27   & 0.09                            & 
3.95                                     &
74     & 4                               &
61     & 6                               \\[1.8ex] 
1.5    & Ar & Ca                         & 
4.5                                      &    
733    & 7                               & 
76     & 4                               & 
0.30   & 0.10                            & 
4.46                                     &
78     & 2                               &
65     & 5                               \\[1.8ex] 
2.0    & C  & C                          & 
3.5                                      &    
651    & 8                               & 
89     & 4                               & 
0.21   & 0.07                            & 
2.39                                     &
84     & 2                               &
71     & 5                               \\ 
2.0    & Ca & Ca                         & 
4.5                                      &    
685    & 9                               & 
86     & 3                               & 
0.30   & 0.08                            & 
0.96                                     &
86     & 2                               &
72     & 5                               \\ 
1.9    & Ni & Ni                         & 
4.5                                      &    
707    & 9                               & 
87     & 4                               & 
0.43   & 0.16                            & 
8.06                                     &
88     & 4                               &
75     & 6                               \\ 
\end{tabular}
\end{table}

\end{document}